\documentclass[%
 reprint,
superscriptaddress,
nofootinbib,
 amsmath,amssymb,
 aps,
 prd,
floatfix,
]{revtex4-2}
\usepackage{tabularx}
\usepackage[table]{xcolor}
\definecolor{tableShade}{gray}{0.9}
\usepackage{graphicx}
\usepackage{dcolumn}
\usepackage{bm}
\usepackage[acronym]{glossaries}

\preto{\abstractkeywords}{\nolinenumbers}
\usepackage{enumitem}
\usepackage{amsmath}
\usepackage[normalem]{ulem}
\usepackage{dcolumn}
\usepackage{tabularx}
\usepackage{multirow}
\usepackage{booktabs}
\usepackage{orcidlink}
\usepackage{import}
\usepackage{amsmath}
\usepackage{amssymb}
\usepackage{graphicx}
\usepackage[normalem]{ulem}
\usepackage{dcolumn}
\usepackage{afterpage}
\usepackage{multirow}
\usepackage{booktabs}
\usepackage{hhline}
\usepackage{makecell}
\usepackage{array}
\usepackage{float}
\usepackage{soul}
\usepackage{comment}
\usepackage{bm}
\usepackage{hyperref}
\hypersetup{
    colorlinks=true,
    linkcolor=blue,
    filecolor=magenta,      
    urlcolor=blue,
    citecolor=violet
}
\usepackage{longtable}

\newcommand{\beq}{\begin{equation}}
\newcommand{\eeq}{\end{equation}}
\newcommand{\bea}{\begin{eqnarray}}
\newcommand{\eea}{\end{eqnarray}}

\newcommand{\blipp}{{\tt BLIP}}
\newcommand{\blip}{{\tt BLIP} }
\newcommand{\blips}{{\tt BLIP}'s\ }
\newcommand{\blipt}{{\tt BLIP 2.0} }
\newcommand{\blippt}{{\tt BLIP 2.0}}

\newcommand{\submodel}{{\tt submodel} }
\newcommand{\model}{{\tt Model} }
\newcommand{\inj}{{\tt Injection} }

\begin{document}

\title{Flexible Spectral Separation of Multiple Isotropic and Anisotropic Stochastic Gravitational Wave Backgrounds in LISA}

\author{Alexander W. Criswell\,\orcidlink{0000-0002-9225-7756}}\email{alexander.criswell@vanderbilt.edu}
\affiliation{Department of Physics \& Astronomy, Vanderbilt University, Nashville, TN 37235, USA}
\affiliation{Department of Life and Physical Sciences, Fisk University, Nashville, TN 37208}
\affiliation{Minnesota Institute for Astrophysics, University of Minnesota, Minneapolis, MN 55455, USA}
\affiliation{School of Physics and Astronomy, University of Minnesota, Minneapolis, MN 55455, USA}

\author{Sharan Banagiri\,\orcidlink{0000-0001-7852-7484}}
\affiliation{Center for Interdisciplinary Exploration and Research in Astrophysics (CIERA), Northwestern University, Evanston, IL 60201, USA}
\affiliation{School of Physics and Astronomy, Monash University, VIC 3800, Australia}
\affiliation{OzGrav: The ARC Centre of Excellence for Gravitational Wave Discovery, Clayton, VIC 3800, Australia}

\author{Jessica Lawrence\,\orcidlink{0000-0003-1222-0433}}
\affiliation{Department of Physics, Texas Tech University, Lubbock, TX 79409, USA}

\author{Levi Schult\,\orcidlink{0000-0001-6425-7807}}
\affiliation{Department of Physics \& Astronomy, Vanderbilt University, Nashville, TN 37235, USA}

\author{Steven Rieck\,\orcidlink{0009-0006-0978-7892}}
\affiliation{School of Physics and Astronomy, University of Minnesota, Minneapolis, MN 55455, USA}
\affiliation{Department of Physics, University of Cincinnati, Cincinnati, OH 45221, USA}

\author{Stephen R. Taylor\,\orcidlink{0000-0001-8217-1599}}
\affiliation{Department of Physics \& Astronomy, Vanderbilt University, Nashville, TN 37235, USA}

\author{Vuk Mandic\,\orcidlink{0000-0001-6333-8621}}
\affiliation{Minnesota Institute for Astrophysics, University of Minnesota, Minneapolis, MN 55455, USA}
\affiliation{School of Physics and Astronomy, University of Minnesota, Minneapolis, MN 55455, USA}

\date{\today}



\begin{abstract}

The Laser Interferometer Space Antenna (LISA) will observe mHz gravitational waves from a wide variety of astrophysical sources. Of these, some will be characterizable as individual deterministic signals; the remainder will overlap to create astrophysical confusion noise. These sources of confusion noise are known as stochastic gravitational wave backgrounds (SGWBs). LISA data is expected to include several such astrophysical SGWBs, including the notable Galactic binary foreground, SGWBs from white dwarf binary populations in satellite galaxies of the Milky Way, and the SGWB from extragalactic stellar-origin binary black holes far from merger. To characterize these astrophysical signals and attempt to seek out possible underlying backgrounds of cosmological origin, it will be necessary to separate the contribution of each SGWB from that of the others. Crucially, several of these SGWBs are expected to be highly anisotropic on the sky, providing a powerful tool for spectral separation. To this end, we present {\tt BLIP 2.0}: a flexible, GPU-accelerated framework for simulation and Bayesian analysis of arbitrary combinations of isotropic and anisotropic SGWBs. We leverage these capabilities to demonstrate for the first time spectral separation of the Galactic foreground, the Large Magellanic Cloud SGWB, and the SGWB from extragalactic stellar-origin binaries, and show a proof-of-concept for placing upper limits on the detection of an underlying isotropic cosmological SGWB in the presence of multiple astrophysical foregrounds.

\end{abstract}

\maketitle

\section{Introduction}\label{sec:intro}
The Laser Interferometer Space Antenna \citep[LISA; ][]{amaro-seoane_laser_2017a,colpi_lisa_2024} is a joint ESA-NASA mission with a current launch date of 2035. It will be a spaceborne interferometric gravitational wave (GW) observatory in the mHz band, formed by a triangular constellation of three spacecraft 2.5 million km apart in a trailing (or leading) Earth orbit. As it occupies an entirely different frequency band to current-day GW observatories, LISA will probe a vast array of astrophysical phenomena as-of-yet unobserved in GWs. These include massive black hole binary (MBHB) mergers \citep{sesana_gravitational_2005}, extreme mass-ratio inspirals (EMRIs;\citep{hils_gradual_1995, sigurdsson_capture_1997,amaro-seoane_intermediate_2007b}), stellar-origin compact binaries far from merger \citep{cutler_what_2019,gerosa_multiband_2019, sesana_prospects_2016, seto_how_2022,babak_stochastic_2023a}, tens of millions of Galactic and extragalactic double white dwarf binaries (DWDs) \citep{nelemans_gravitational_2001,edlund_white_2005a,ruiter_lisa_2010}, and more; see \citet{amaro-seoane_astrophysics_2023a} for a review.

In principle, any of the astrophysical GW sources discussed above could give rise to a stochastic gravitational wave background (SGWB) in LISA. In contrast to individually-resolvable, bright signals, SGWBs are formed from the superposition of many faint GW sources which collectively produce a GW confusion ``noise". SGWBs have been searched for in terrestrial gravitational-wave observatories \citep{abbott_upper_2017a,abbott_search_2019b,abbott_search_2021a,abbott_upper_2021a}, and evidence for a nHz SGWB has been observed by pulsar timing arrays (PTAs) \citep{agazie_nanograv_2023e, agazie_nanograv_2023h,agazie_nanograv_2023k,antoniadis_second_2023a,antoniadis_second_2023d,reardon_gravitationalwave_2023a,reardon_search_2023a,zic_parkes_2023a,xu_searching_2023a,miles_meerkat_2025}. Depending on the nature of the astrophysical or cosmological phenomena underpinning a given SGWB, the resulting signal can have different spectral characteristics, spatial morphology (i.e. anisotropy or lack thereof), and polarization content.\footnote{This work focuses on the spectral and spatial distributions of LISA SGWBs, and leaves consideration of polarization as a compelling direction of future work.} As we will see, leveraging these facets of SGWB signals can prove useful for distinguishing the respective contributions of several such signals --- a feat known as spectral separation. The ability to perform this kind of spectral separation will be crucial for SGWB science with LISA, as it is expected that the LISA datastream will contain several such signals.

Foremost among these is the stochastic GW contribution from every unresolved mHz DWD system in the Milky Way (MW), known as the Galactic foreground \citep{nelemans_gravitational_2001,edlund_white_2005a,ruiter_lisa_2010}.\footnote{So named due to its prominence above even the LISA instrumental noise. For clarity and conciseness, we take the term ``SGWB" to be inclusive of the Galactic foreground throughout this work, despite its ``foreground" status.} The Galactic foreground dominates the LISA band below a few mHz, where it rapidly truncates due to the subtraction of resolvable sources, source discreteness, and stellar evolutionary effects \citep{nelemans_gravitational_2001,edlund_white_2005a,ruiter_lisa_2010,cornish_galactic_2017a,korol_observationally_2021}. Due to its large amplitude, accurate characterization of the foreground signal is of paramount importance for detection and characterization of not just other SGWBs, but resolvable signals as well. In addition to its characteristic truncated spectrum, the Galactic foreground is highly anisotropic, tracing the spatial distribution of the Milky Way's DWD population on the sky. This not only proves to be a valuable asset for spectral separation between the foreground and underlying SGWBs, but can also provide us with valuable insights into our Galaxy \citep[e.g.,][]{breivik_constraining_2020c,zhang_constraining_2024a,adams_astrophysical_2012a}.

The Milky Way is not, however, the only significant nearby population of DWD systems. \citet{rieck_stochastic_2024} show that the unresolved DWD systems in the Large Magellanic Cloud (LMC) produce a significant anisotropic SGWB (ASGWB) in LISA. Other MW satellites may contribute their own, lower-amplitude ASGWBs \citep{pozzoli_cyclostationary_2024}.

LISA is also expected to observe the astrophysical SGWB from extragalactic stellar-origin compact binaries. This background consists of binaries that will eventually merge in terrestrial GW detectors like that of the LIGO-Virgo-KAGRA (LVK) Collaborations \citep{aasi_advanced_2015c,acernese_advanced_2015a,akutsu_kagra_2019b}. However, these systems are far from merger while still in the mHz; as such, this SGWB is expected to possess an $\alpha=2/3$ power law spectrum \citep{phinney_practical_2001}.\footnote{$\alpha=2/3$ in $\Omega_{\rm GW}$; alternatively, $-7/3$ in PSD or $-2/3$ in characteristic strain.} As the dominant contribution is expected to be from binary black hole mergers \citep[e.g.,][]{abbott_upper_2021a}, this signal is often referred to as the stellar-origin binary black hole (SOBBH) background; we will adopt this terminology moving forward. Current LVK rate measurements of the binary black hole (BBH) population \citep[e.g.,][]{abbott_gwtc3_2023a} can be used to predict the amplitude of the SOBBH SGWB. For a $\alpha=2/3$ power law with $f_{\mathrm{ref}}=25$ Hz, the SOBBH SGWB amplitude is expected to be $\Omega_{\mathrm{ref}}=7.3^{+4.5}_{-4.6}\times10^{-10}$, as estimated for the LISA band by \citet{babak_stochastic_2023a}; this estimate is consistent with those of other recent studies \citep[e.g.,][]{chen_stochastic_2019, cusin_properties_2019a, perigois_startrack_2021,abbott_upper_2021a,lewicki_impact_2023}. The SOBBH SGWB will likely manifest as an isotropic SGWB (ISGWB) as seen in LISA.

Beyond these likely signals, there remain many other, more uncertain prospects. As discussed earlier, any resolved LISA source can have a corresponding background arising from its unresolved counterparts. However, whether the rate of the source in question is sufficient to generate a SGWB (let alone a detectable SGWB) in the LISA band is a different question. It is possible that LISA may observe a SGWB from EMRIs \citep{bonetti_gravitational_2020b, naoz_enhanced_2023b,pozzoli_computation_2023a,piarulli_test_2024a,xuan_stochastic_2024}, MBHB mergers, or other populations of unresolved DWD systems beyond the MW and its satellites \citep{hofman_uncertainty_2024, staelens_likelihood_2024}. Finally, a broad array of proposed cosmological SGWBs (i.e., inherently stochastic fields from the early universe) may be detectable (or at least able-to-be-constrained) by LISA; see \citet{auclair_cosmology_2023a} for a review.

\subsection{LISA Data Analysis}\label{sec:intro_data_analysis}
This plethora of sources carries simultaneously both great scientific potential and immense technical difficulty. In contrast to current terrestrial GW observatories which observe individual GW signals interspersed among long stretches of detector noise, the LISA datastream will instead be \textit{signal-dominated} due to the sheer number of GW sources emitting constantly in the LISA band. In short, we will not know what LISA data looks like without GW signals in it, which lends considerable difficulty to the task of disambiguating GW signals from instrumental noise. This aspect of the LISA data analysis problem has motivated a ``fit everything, everywhere, all at once" approach, dubbed the Global Fit \citep{cornish_lisa_2005, crowder_solution_2007,littenberg_detection_2011a,littenberg_global_2020,littenberg_prototype_2023}. 

The majority of current Global Fit prototypes \citep{littenberg_prototype_2023,katz_efficient_2024,deng_modular_2025} --- with the exception of the maximum-likelihood approach of \citet{strub_global_2024} --- take the form of $\mathcal{O}(100,000)$-parameter transdimensional Bayesian analyses with a blocked structure. Their transdimensional nature allows for inference from the data of not just individual source parameters, but also the presence/absence and overall number of GW signals in the data. This is accomplished through Reverse Jump Markov Chain Monte Carlo (RJMCMC) sampling; such Global Fits are therefore necessarily Bayesian in nature. The blocked nature of the Global Fit structure comprises a high-level Gibbs sampler which allows for each source category (DWDs, MBHBs, EMRIs, SGWBs, etc.) to be treated by a purpose-made module, with information passing from module to module at every cycle of the sampler. Analyses which seek to constrain a single LISA source category must themselves be Bayesian as well, if they are to be integrated into a Global Fit setting of this kind.

Finally, due to otherwise-excessive amounts of laser phase noise, LISA data analysis will be performed using Time-Delay Interferometry (TDI) channels constructed virtually from the individual Doppler-tracking data of each LISA link \citep{tinto_timedelay_2020}. The analyses considered in this work use the AET channels \citep{tinto_timedelay_2020,tinto_secondgeneration_2023}.

\subsection{LISA Multi-SGWB Analyses}\label{sec:intro_sgwb_analyses}
Building on previous single-signal works \citep{contaldi_maximum_2020,banagiri_mapping_2021a,caprini_reconstructing_2019,ungarelli_studying_2001,kudoh_probing_2005,taruya_probing_2005,renzini_mapping_2018,cornish_detecting_2001,cornish_space_2001a,adams_discriminating_2010a}, a number of prototype analyses have been developed to analyze multiple SGWB signals in the LISA datastream. These include: SGWB searches in Global Fit residuals, following characterization of the foreground by the Global Fit itself \citep{rosati_prototype_2024}; two-component spectral separation of the Galactic foreground and a single background signal \citep{boileau_spectral_2021a,hindmarsh_recovering_2024,lin_white_2023,mentasti_probing_2024,rieck_stochastic_2024}; two-component spectral separation of an astrophysical and cosmological background \citep{boileau_spectral_2021a}; three-component separation of a simplified Galactic foreground (i.e. sans treatment of anisotropy or its induced time-modulation), the SOBBH astrophysical background, and an underlying cosmological signal \citep{pieroni_foreground_2020a,flauger_improved_2021a,braglia_gravitational_2024a,liang_unveiling_2024}. These analyses, while promising in many respects, largely neglect to model anisotropies in either the Galactic foreground or potential underlying anisotropic signals.\footnote{The exceptions being \citet{rieck_stochastic_2024}, which used \blip to analyze the LMC and employed an early prototype of the framework presented here; \citet{hindmarsh_recovering_2024}, which considers a cosmological signal and a time-modulated Galactic foreground; and \citet{pozzoli_cyclostationary_2024}, which leverages the cyclostationary statistics of the induced time-modulation of anisotropic SGWBs to separate the contributions from simple models of the Galactic foreground and the LMC.} In doing so, they discard crucial information that can not only be used to characterize the MW itself \citep[e.g.,][]{breivik_constraining_2020c}, but can possess significant utility when attempting to separate the anisotropic Galactic foreground from underlying isotropic and anisotropic SGWBs. There is a pressing need in the field for a SGWB analysis for LISA that is 1) Bayesian, so as to be compatible with the Global Fit; 2) capable of characterizing the anisotropic nature of the Galactic foreground and other anisotropic SGWBs like that of the Large Magellanic Cloud; 3) built for flexible spectral separation of multiple SGWBs in the LISA datastream; and 4) designed with eventual transdimensional RJMCMC SGWB analyses in mind.

In this work, we build on the Bayesian LISA Inference Package \citep[\blipp;][]{banagiri_mapping_2021a} to develop a flexible simultaneous inference formalism for both isotropic and anisotropic SGWBs in LISA. We leverage the anisotropic analysis capabilities of \blip detailed in \citet{banagiri_mapping_2021a,criswell_templated_2025} and fully refactor the \blip codebase to implement a fully modular structure --- \blipt --- that not only enables flexible construction of spectral separation models for any number of isotropic and/or anisotropic SGWBs alongside simulation of the same, but has also been designed with an eventual transdimensional RJMCMC analysis in mind. The statistical formalism of this spectral separation infrastructure is developed in \S\ref{sec:si_formalism}. Its implementation as \blipt is described in \S\ref{sec:blip2}. We provide proofs-of-concept for the \blipt infrastructure via analyses of simulated LISA multi-SGWB data, including the Galactic foreground, the ASGWB from white dwarf binaries in the LMC, the isotropic SGWB from extragalactic stellar-origin binary black holes, and an underlying isotropic cosmological SGWB. The simulation procedure is discussed in \S\ref{sec:sims}, and the results of the corresponding \blipt spectral separation analyses are shown in \S\ref{sec:results}. Discussion of the results and directions of future work are explored in \S\ref{sec:discussion_conclusions}.

\section{Simultaneous Inference Formalism}\label{sec:si_formalism}
\subsection{BLIP Single-SGWB Analyses}\label{sec:si_single}
We present in brief the \blip likelihood for a single signal, as discussed in \citet{banagiri_mapping_2021a,criswell_templated_2025}; these derivations in turn built on those of previous works \citep{cornish_detecting_2001,cornish_space_2001a,adams_discriminating_2010a,romano_detection_2017}. For further details, refer to \citet{banagiri_mapping_2021a,criswell_templated_2025}. 

\blip analyses divide the LISA time-domain datastream into segments of duration $T_{\mathrm{seg}}=10^5$ s, and perform a short-time Fourier transform with Hann windowing on each segment. Each segment is assumed independent and combined in a variant multidimensional complex Gaussian likelihood \citep{adams_discriminating_2010a}:

\begin{widetext}
\begin{equation}\label{eq:blip_likelihood}
    \mathcal{L}(d|\vec\theta) = \prod_{f,t}\frac{1}{2\pi T_{\mathrm{seg}}\det\left(C_{IJ}(f,t|\vec\theta)\right)}
    \times\exp\left[-\frac{2}{T_{\mathrm{seg}}}\sum_{I,J}\tilde{d}_I^*(f,t)\left(C(f,t|\vec\theta)\right)^{-1}_{IJ}\tilde{d}_J(f,t)\right],
\end{equation}
\end{widetext}
where $C_{IJ}(f,t|\vec\theta)$ is the time- and frequency-dependent channel covariance matrix as a function of some model parameters $\vec{\theta}$. $f$ and $t$ here index over the $f^{th}$ frequency bin/$t^{th}$ time segment, respectively. $I$ and $J$ index over LISA's three TDI channels (XYZ or AET). Unless otherwise noted, we use the AET channels. $\tilde{d}_I$ denotes the Fourier-domain data in channel $I$.

For a single SGWB in the presence of noise, we can write
\begin{equation}
    C_{IJ}(f,t) = S^n_{IJ}(f) + S^{\mathrm{GW}}_{IJ}(f,t),
\end{equation}
where $S^n_{IJ}$ and $S^{\mathrm{GW}}_{IJ}$ denote the noise and SGWB cross-/autopower spectral densities (PSDs), respectively. The former depends solely on the LISA instrument; we detail the simple LISA noise model used in this work in \S\ref{sec:blip2_spec_noise}. The latter, however, accounts for both the intrinsic nature of the SGWB in question, as well as its interaction with the LISA detector response. In general, the SGWB power can be described in terms of its spectral and spatial distribution as $S_{\mathrm{GW}}(f,\mathbf{n})$, where $\mathbf{n}$ denotes direction on the sky, i.e., from LISA to the GW origin. We assume that the SGWB spectrum $S_{\mathrm{GW}}(f)$ and spatial distribution $\mathcal{P}(\mathbf{n})$ are separable; that is,
\begin{equation}\label{eq:si_separable}
    S_{\mathrm{GW}}(f,\mathbf{n}) = S_{\mathrm{GW}}(f)\mathcal{P}(\mathbf{n}),
\end{equation}
where $\mathcal{P}(\mathbf{n})$ is normalized such that
\begin{equation}\label{eq:si_single_Pofn_norm}
    \int\mathcal{P}(\mathbf{n})d^2\mathbf{n} = 1.
\end{equation}
For an isotropic SGWB, $\mathcal{P}(\mathbf{n}) = 1/4\pi$. \blip is capable of modelling anisotropic SGWBs in terms of a spherical harmonic expansion \citep{banagiri_mapping_2021a} or a pixelated skymap \citep{criswell_templated_2025}. Depending on the nature of the an/isotropic model, the SGWB signal is convolved with the appropriate time/frequency/directionally-dependent LISA response $\mathcal{R}_{IJ}(f,t,\mathbf{n})$ such that
\begin{equation}\label{eq:si_skymap_convolution}
    S^{\mathrm{GW}}_{IJ}(f,t) = \int\mathcal{P}(\mathbf{n})\,\mathcal{R}_{IJ}(f,t,\mathbf{n})\,d^2\mathbf{n} \times S_{\mathrm{GW}}(f).
\end{equation}
For full derivations, we refer the reader to \citet{banagiri_mapping_2021a} for the spherical harmonic case, \citet{criswell_templated_2025} for the pixel-basis case, and \citet{cornish_detecting_2001,cornish_space_2001a,adams_discriminating_2010a,romano_detection_2017} for the general case.

\subsection{Extension to Multiple Signals with Diverse Morphologies}\label{sec:si_multi}
We now consider the case of simultaneous inference of multiple SGWBs in the presence of instrumental noise, allowing for diverse SGWB spectral and spatial morphologies. The likelihood for this case is a straightforward extension of the single-signal likelihood of Eq.~\eqref{eq:blip_likelihood}. As the covariance contributions of multiple (uncorrelated) signals in each channel are additive, we can write the likelihood for simultaneous inference of $N$ signal/noise (sub)models $M_{1...N}$ as
\begin{widetext}
\begin{equation}\label{eq:blip_si_likelihood}
    \mathcal{L}_{\mathrm{SI}}(d|\vec{\theta}) = \prod_{t,f}\left[ \frac{1}{2\pi T_{\mathrm{seg}}\det\left(C_{IJ}(t,f|M_{1...N};\vec\theta)\right)}\times\exp\Biggl[-\frac{2}{T_{\mathrm{seg}}}\sum_{I,J}\tilde{d}_I^*(t,f)\left(C_{IJ}(t,f|M_{1...N};\vec\theta)\right)^{-1}\tilde{d}_J(t,f)\Biggr]\right],
\end{equation}
\end{widetext}
with $\vec\theta = \{\vec\theta_k\}$ for $k\in[1...N]$ such that $\vec\theta_k$ are the parameters of signal/noise (sub)model $M_k$. ``$\mathrm{SI}$" denotes the simultaneous inference case. The cumulative model covariance is defined as
\begin{equation}\label{eq:blip_si_cov}
    C_{IJ}(t,f|M_{1...N};\vec\theta) = \sum_{k=1}^{N} C_{IJ}^{M_k}(t,f|\vec\theta_k),
\end{equation}
Note that only the spatial and spectral distributions of \textit{individual} ASGWBs are assumed to be separable; the overall (i.e., cumulative) spatial and spectral distributions of SGWB power are no longer separable as defined by Eq.~\eqref{eq:si_separable}.

\subsection{Example: Covariance for Three Components}\label{sec:blip_si_cov_example}
For example, consider a simultaneous inference model consisting of the LISA instrumental noise and two SGWB signals: an isotropic power law --- e.g., a cosmological or SOBBH ISGWB --- and an anisotropic tanh-truncated power law with a spherical harmonic parameterized spatial model, motivated by simplified models of the Galactic foreground; see \S\ref{sec:blip2_models} for detailed discussion of these models. The overall parameter vector is then 
\begin{widetext}
\begin{equation}
\begin{split}
    \vec\theta &= \bigl\{\vec\theta_n,\vec\theta_{\mathrm{PL}},\vec\theta_{\mathrm{TPL}}\bigr\}\\
    &=\bigl\{\vec\theta_n,\vec\theta_{\mathrm{PL}},\{\vec\theta_{\mathcal{S},\mathrm{TPL}},\vec\theta_{f,\mathrm{TPL}}\}\bigr\}\\
    &=\bigl\{\{N_a,N_p\},\{\alpha_{\mathrm{PL}},\Omega_{\mathrm{ref, PL}}\},\{\vec b_{\ell m},\alpha_{\mathrm{TPL}},\Omega_{\mathrm{ref, TPL}},f_{\mathrm{break}},f_{\mathrm{scale}}\}\bigr\},
\end{split}
\end{equation}
\end{widetext}
where PL and TPL refer to the power law and truncated power law (sub)models, respectively. A given channel's model covariance contribution is then (using TDI X channel auto-power as an example):\footnote{Note that $\mathcal{R}_{XX}(t,f)$ in Eq.~\eqref{eq:example_multi_xx} is the integral of $\mathcal{R}_{XX}(t,f,\mathbf{n})$ for the isotropic case; see Eq.~\eqref{eq:si_skymap_convolution} and surrounding discussion.}
\begin{widetext}
\begin{equation}
\begin{split}
    S_{XX}(t,f|M_{1...N};\vec\theta) =&\ S_{XX}^{n,M_n}(t,f|\vec\theta_n) \\
    &+ \mathcal{R}_{XX}(t,f) S_{\mathrm{GW}}^{M_{\mathrm{PL}}}(f|\vec\theta_{\mathrm{PL}}) \\
    &+ \int\mathcal{R}_{XX}(t,f,\mathbf{n})\mathcal{P}(\mathbf{n}|\vec\theta_{\mathcal{S},\mathrm{TPL}})d^2\mathbf{n}\times S_{\mathrm{GW}}^{M_{\mathrm{TPL}}}(f|\vec\theta_{f,\mathrm{TPL}}).
\end{split}
\label{eq:example_multi_xx}
\end{equation}
\end{widetext}
This result generalizes to all cross- and autopower channels to construct the full covariance matrix.

\section{\blipt: a Flexible Simultaneous Inference Framework}\label{sec:blip2}
As elegant in its simplicity as the formulation described in the previous section may be, its practical implementation in the general case is deceptively complex. While it is possible to implement individual cases of interest in the original \blip infrastructure \citep[as seen in][]{rieck_stochastic_2024}, doing so repeatedly requires considerable duplicated effort and is ultimately far too inefficient for general use. As discussed in \S\ref{sec:intro}, there exist a wide variety of potential LISA SGWBs; these span varied spectral and spatial morphologies which can of course be described by any number of potential parametric (or even nonparametric) models. Moreover, flexibility in choice of both spectral and spatial models is highly valuable in a Global Fit context, where the RJMCMC nature of the Global Fit could in principle be used to perform SGWB model selection during the analysis. The ideal solution is then a framework in which the simultaneous inference formalism of \S\ref{sec:si_formalism} is flexibly implemented for an arbitrary number and combination of signal models --- as well as one in which implementing new signal models is simple and straightforward.

\begin{figure*}
    \centering
    \includegraphics[width=1.0\linewidth]{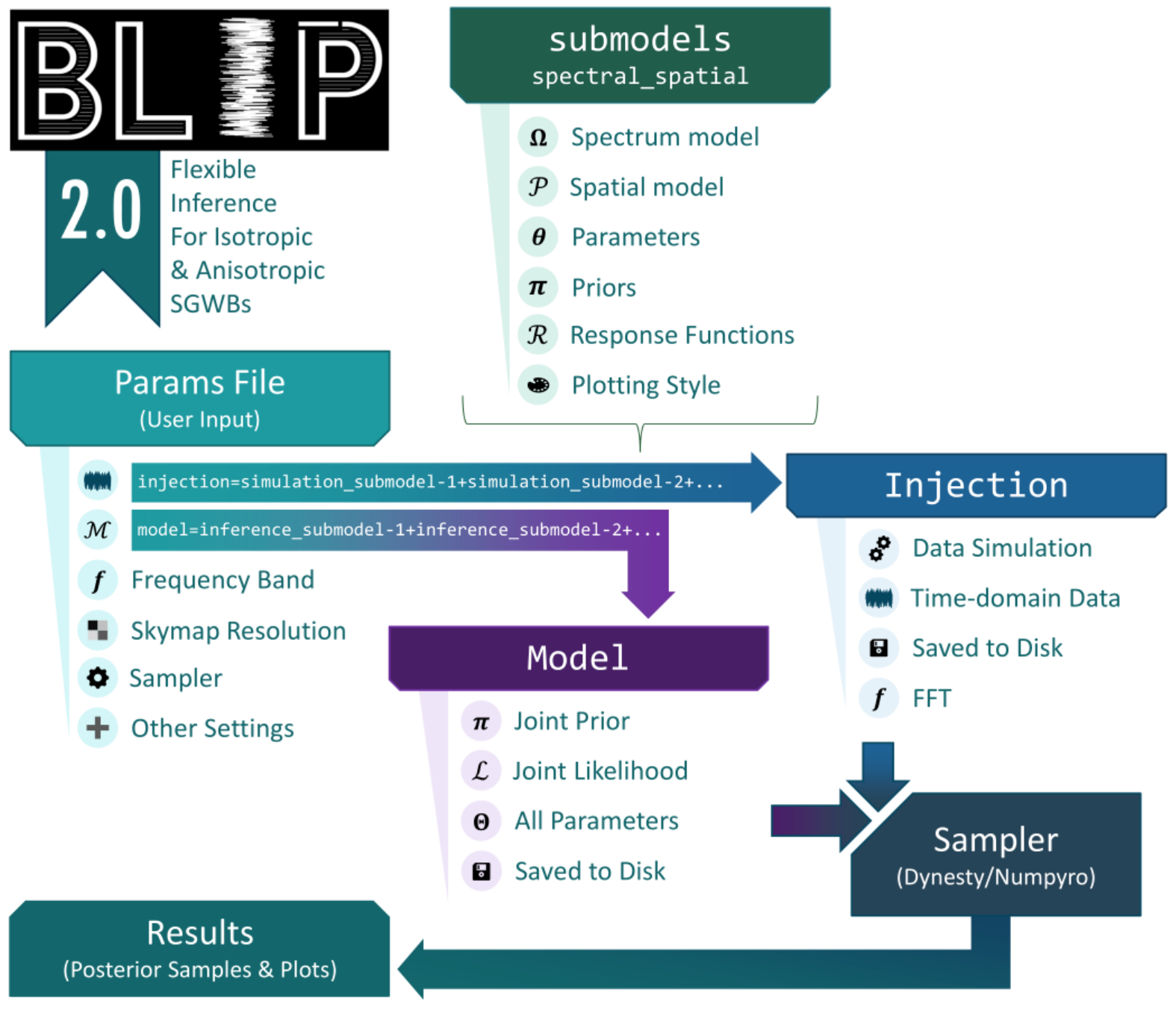}
    \caption{Schematic diagram of the \blip 2.0 code structure.}
    \label{fig:blip_code_diagram}
\end{figure*}

To this end, the \blip codebase has been refactored from the ground up to enable full flexibility in both simulation and analysis of isotropic and anisotropic SGWBs. This is accomplished through a fundamentally modular design, dubbed \blipt. All simulation and analysis routines have been rewritten to interact with a unified \submodel class, along with the \inj and \model classes that marshal one or more {\tt submodel}s for use in simulation or analysis, respectively. Each instantiated \submodel (with the exception of the LISA instrumental noise) consists of a spectral and spatial model, and carries with it all relevant information and functions related to one SGWB signal: its spectrum, its spatial distribution, the LISA response to it, its model parameters and prior distributions (in the case of an inference {\tt submodel}), aesthetic details and names for plotting purposes, and so on. The \submodel for the LISA instrumental noise is structured in a similar fashion, but does not contain a response function or spatial distribution. The \model class constructs the overall likelihood and prior distribution from a collection of {\tt submodel}s, and is the primary object (along with the data) that is passed to the sampler. For a single signal + instrumental noise, this likelihood is the one discussed in \S\ref{sec:si_single}. In the case of multiple signals in the presence of instrumental noise, the \model class implements the simultaneous inference likelihood discussed in \S\ref{sec:si_multi}. The \inj class performs a similar function for simulations, combining the contributions from multiple {\tt submodel}s. Refer to Fig.~\ref{fig:blip_code_diagram} for a visual outline of the \blip 2.0 infrastructure.

Within this new framework, to add a new spectral or spatial model to {\tt BLIP}, all one must do is provide a name, a function governing its spectral or spatial distribution, the parameters for that function, its prior distribution, and --- in the case of a spatial model --- designate which of the LISA response routines it should connect to (i.e., isotropic responses for an ISGWB signal, anisotropic responses for an ASGWB). Anisotropic spatial models can be represented in either the pixel or spherical harmonic basis. Any combination of spectral and spatial models can be combined to form a {\tt submodel}; this is accomplished easily by the end-user via a string within the \blip input (``params") file. For example, to simulate and analyze a power law ISGWB in the presence of LISA instrumental noise, one simply enters 
\begin{equation*}
    \text{{\tt noise+powerlaw\_isgwb}}
\end{equation*}
for both the \model and \inj items in the params file. These specifications need not be matched, and --- as will be demonstrated in \S\ref{sec:sims}-\ref{sec:results} --- can be formed of any arbitrary combination of {\tt submodel}s. Relevant information for each {\tt submodel}, such as the true parameter values for a simulated signal, can be passed by the user to \blip via dictionaries in the params file where each submodel string operates as a key. This framework has enabled the implementation of a large number of new spectral and spatial models within {\tt BLIP}; these models are detailed in \S\ref{sec:blip2_models}. 

Additionally, the {\tt Model}/{\tt Injection}/{\tt submodel} structure not only allows relevant information to be seamlessly carried around \blip itself for simulation, analysis, and plotting of results, but also enables data availability by saving the \model and \inj objects (and all {\tt submodel}s therein) alongside the posterior samples, plots, and so on. As a result, all information pertaining to a \blip simulation/analysis run can be easily accessed after it is complete and uploaded to open data repositories such as Zenodo.

\subsection{Dynamic Assembly of Simultaneous Inference Models in \blipt}\label{sec:blip_si_implementation}
The modular\  \model/ \submodel infrastructure of \blip 2.0 is uniquely suited to support simultaneous inference analyses. The \blip \model object's likelihood and covariance routines are designed to flexibly implement the simultaneous inference formalism described above with arbitrary combinations of {\tt submodel}s, across any number of signals.\footnote{With the following disclaimers: subject to available computing power; poorly- or over-specified models may not converge. Just because one \textit{can} create a model with an arbitrary number of signals, doesn't necessarily mean one \textit{should}. As always, discretion is key.} To the author's knowledge at time of writing, \blip is the only data analysis software for LISA with the capability to flexibly characterize arbitrary combinations of isotropic and anisotropic SGWBs. 

To elucidate how this is accomplished in practice, consider again the example of \S\ref{sec:blip_si_cov_example}. To perform an analysis using that simultaneous inference model --- i.e., a power-law ISGWB paired with a truncated-power law ASGWB described by the spherical harmonic spatial model --- we simply set
\begin{equation*}
    \text{{\tt model=noise+powerlaw\_isgwb+truncatedpowerlaw\_sph}}.
\end{equation*}
\blip will then construct a \model from these {\tt submodel}s by computing the appropriate LISA response functions for each,\footnote{With the exception of the LISA detector noise, which draws on its own set of unique routines and does not require the calculation of a response function.} specifying the form of each {\tt submodel}'s covariance contribution (and its dependence on that {\tt submodel}'s spatial and spectral parameters), and combining those covariance contributions within the formalism of \S\ref{sec:si_multi}. Throughout this process, \blip automatically handles all the requisite miscellanea to ensure that the sampler treats each {\tt submodel}'s prior draws appropriately, parameter ordering is maintained throughout the code, duplicate model specifications are clearly differentiated, and so on. The model string given above can comprise of any number of ISGWB and ASGWB submodels. Duplicate submodels can be described by, e.g.,
\begin{equation*}
    \text{{\tt model=noise+powerlaw\_isgwb-1+powerlaw\_isgwb-2}}.
\end{equation*}
Every included submodel is ascribed a unique identifier, labeled as such on all plots produced at the end of an analysis run, and color-coded accordingly. This structure allows \blip to seamlessly support almost any simultaneous inference analysis of interest, as will be seen in \S\ref{sec:results}.

\subsection{Currently Available Submodels}\label{sec:blip2_models}
Table~\ref{tab:models} summarizes the currently available spectral and spatial submodels within \blippt. The implementation within \blip of many of these have been described elsewhere; see the listed references where relevant. The SGWB spectral models are parameterized in terms of the dimensionless GW energy density $\Omega_{\mathrm{GW}}$, but they are ultimately transformed into the (equivalent) strain PSD $S_{\rm GW}$. These quantities are related via
\begin{equation}\label{eq:omegaf_sgw}
    \Omega_{\mathrm{GW}}(f) = \frac{2\pi^2}{3H_0^2}f^3S_{\mathrm{GW}}(f).
\end{equation}
Models not already described elsewhere are elaborated on below. Each model can be used flexibly for simulation and/or inference, and the spatial models can be implemented in either the pixel- or spherical harmonic basis. Only the spherical harmonic spatial model currently allows for inference of spatial parameters when used as an inference model; pixel-basis spatial models use the specified distribution as a fixed spatial template as described at length in \citet{criswell_templated_2025}. Enabling parameter inference for some of these latter spatial models is an ongoing effort beyond the scope of this work.

{\setlength{\extrarowheight}{5pt}%
\begin{table*}
    \centering
    \begin{tabular}{|c||c|c|c|c|}
    \hline
         \textbf{Name}&  \textbf{Type}&  \textbf{Parameters}&  \textbf{Formula}&  \textbf{Reference} \\
         \hline
         Power Law& Spectral & $\Omega_{\mathrm{ref}},\alpha$ & $\Omega_{\mathrm{GW}}(f) = \Omega_{\mathrm{ref}}\left(\frac{f}{f_{\mathrm{ref}}}\right)^{\alpha}$ & \citet{banagiri_mapping_2021a} \\[5 pt]
         \hline
         \multirow{2}{6em}{\centering Broken Power Law} & \multirow{2}{4em}{\centering Spectral} & \multirow{2}{8em}{\centering $\Omega_{\mathrm{ref}}$, $\alpha_1$, $\alpha_2$, $f_{\mathrm{break}}$, $\delta$} & \multirow{2}{22em}{\centering $\Omega_{\mathrm{GW}}(f) = \Omega_{\mathrm{ref}}\left(\frac{f}{f_{\mathrm{ref}}}\right)^{\alpha_1} \bigg(1 + \left(\frac{f}{f_{\mathrm{break}}}\right)^{\frac{1}{\delta}}\bigg)^{\delta(\alpha_1 - \alpha_2)}$} & \S\ref{sec:blip2_models_bpl} \\
          &  &  &  & \\[5pt]
         \hline
         \multirow{2}{6em}{\centering Truncated Power Law} & \multirow{2}{4em}{\centering Spectral} & \multirow{2}{8em}{\centering $\Omega_{\mathrm{ref}}$, $\alpha$, $f_{\mathrm{cut}}$, $f_{\mathrm{scale}}$} & \multirow{2}{22em}{\centering $\Omega_{\mathrm{GW}}(f) = \frac{1}{2}\,\Omega_{\mathrm{ref}}\left(\frac{f}{f_{\mathrm{ref}}}\right)^{\alpha} \bigg(1 + \tanh\left(\frac{f_{\mathrm{cut}} - f}{f_{\mathrm{scale}}}\right)\bigg)$} & \citet{criswell_templated_2025} \\
         &  &  &  & \\[5 pt]
         \hline
         \multirow{2}{6em}{\centering Analytic Foreground} & \multirow{2}{4em}{\centering Spectral} & \multirow{2}{8em}{\centering $\Omega_{\mathrm{ref}}$ (see note)} & \multirow{2}{22em}{\centering $\Omega_{\mathrm{GW}}(f)=\Omega_{\mathrm{ref}}\left(\frac{f}{f_{\mathrm{ref}}}\right)^{2/3}e^{-f^{\alpha_{\mathrm{C17}}}-\beta f \sin(\kappa f)}\newline \times\left[1 + \tanh(\gamma (f_k - f))\right]$} & \S\ref{sec:blip2_models_analyticfg} \\
         &  &  &  & \\[5 pt]
         \hline
         \multirow{2}{6em}{\centering Spherical Harmonic} & \multirow{2}{4em}{\centering Spatial} & \multirow{2}{8em}{\centering $\vec{b}_{\ell m}$} & \multirow{2}{22em}{\centering Spherical harmonic expansion of $\sqrt{\mathcal{P}(\mathbf{n})}$; see ref.} & \citet{banagiri_mapping_2021a} \\
         &  &  &  & \\[5 pt]
         \hline
         \multirow{2}{6em}{\centering Analytic Galaxy} & \multirow{2}{4em}{\centering Spatial} & \multirow{2}{8em}{\centering $r_h$, $z_h$} & \multirow{2}{22em}{\centering Pixel-basis template created from a simple bulge + exponential disk density distribution; see ref.} & \citet{criswell_templated_2025} \\
         &  &  &  & \\[5 pt]
         \hline
         \multirow{2}{6em}{\centering Analytic Satellite} & \multirow{2}{4em}{\centering Spatial} & \multirow{2}{8em}{\centering $d$, $r$, RA, DEC} & \multirow{2}{22em}{\centering Pixel-basis template created from a simple spherical density distribution.} & \S\ref{sec:blip2_spatial_satellite} \\
         &  &  &  & \\[5 pt]
         \hline
         \multirow{2}{6em}{\centering Point Source (Radiometer)} & \multirow{2}{4em}{\centering Spatial} & \multirow{2}{8em}{\centering $\{\theta_i,\phi_i\}_{i\in1...N}$} & \multirow{2}{22em}{\centering Pixel-basis template created from a variable number of point sources.} & \S\ref{sec:blip2_spatial_radiometer} \\
         &  &  &  & \\[5 pt]
         \hline
         \multirow{2}{6em}{\centering {\tt popmap}} & \multirow{2}{4em}{\centering Spatial} & \multirow{2}{8em}{\centering \textbf{---}} & \multirow{2}{22em}{\centering Pixel-basis template created from any DWD population synthesis catalogue; see ref.} & \citet{criswell_templated_2025} \\
         &  &  &  & \\[5 pt]
         \hline
    \end{tabular}
    \caption{Summary of current models available in \blipp, both spectral and spatial. For further details, see the listed reference or section; for parameter definitions refer to Table~\ref{tab:parameter_defs}. (\textbf{Note:} the other parameters in the analytic foreground spectral model are fixed for a given observing duration. See \ref{sec:blip2_models_analyticfg} for details.)}
    \label{tab:models}
\end{table*}}

{\setlength{\extrarowheight}{3pt}%
\begin{table*}
    \centering
    \begin{tabular}{|c|c|c|}
    \hline
        Parameter & Definition & Reference \\[3 pt]
         \hline
        $\Omega_{\mathrm{ref}}$ & Amplitude at $f_{\mathrm{ref}}$, in dimensionless GW energy density. & \citet{banagiri_mapping_2021a}\\[3 pt]
        \hline
        $f_{\mathrm{ref}}$ & Reference frequency. 1 mHz unless specified otherwise. & \citet{banagiri_mapping_2021a}\\[3 pt]
        \hline
        $\alpha$ & Power law slope. Subscripts (e.g. $\alpha_1$, $\alpha_2$) denote multiple power laws. & \citet{banagiri_mapping_2021a} \\[3 pt]
         \hline
        $f_{\mathrm{break}}$ & Break frequency for the broken power law model. & \S\ref{sec:blip2_models_bpl}\\[3 pt]
         \hline
        $\delta$ & Smoothing parameter the broken power law model. & \S\ref{sec:blip2_models_bpl}\\[3 pt]
         \hline
        $f_{\mathrm{cut}}$ & Truncation frequency for the truncated power law model. & \citet{criswell_templated_2025}\\[3 pt]
         \hline
        $f_{\mathrm{scale}}$ & Truncation scaling parameter for the truncated power law model. & \citet{criswell_templated_2025}\\[3 pt]
         \hline
         $\vec{b}_{\ell m}$ & Coefficients of the spherical harmonic expansion of $\sqrt{\mathcal{P}(\mathbf{n})}$ & \citet{banagiri_mapping_2021a}\\[3 pt]
         \hline
        $r_h$ & Radial scale height for the analytic Milky Way disk model of \citet{breivik_constraining_2020c}. & \citet{criswell_templated_2025} \\[3 pt]
         \hline
        $z_h$ & Vertical scale height for the analytic Milky Way disk model of \citet{breivik_constraining_2020c}. & \citet{criswell_templated_2025} \\[3 pt]
         \hline
        $d$ & Distance to the center of the analytic MW satellite. & \S\ref{sec:blip2_spatial_satellite}\\[3 pt]
         \hline
        $r$ & Radius of the (assumed spherical) analytic MW satellite. & \S\ref{sec:blip2_spatial_satellite}\\[3 pt]
         \hline
        RA, DEC & Right ascension and declination of the analytic MW satellite. & \S\ref{sec:blip2_spatial_satellite}\\[3 pt]
         \hline
        $\theta_i$, $\phi_i$ & Sky location of the $i^{\mathrm{th}}$ point source, for $i\in1...N$. Can also be given in RA/DEC. & \S\ref{sec:blip2_spatial_radiometer}\\[3 pt]
         \hline
    \end{tabular}
    \caption{Parameter definitions for the models discussed in this work.}
    \label{tab:parameter_defs}
\end{table*}
}

\subsubsection{Broken Power Law (Spectral)}\label{sec:blip2_models_bpl}
The broken power law spectral model,
\begin{widetext}
\begin{equation}\label{eq:blip_bpl}
    \Omega_{\mathrm{GW}}(f) = \Omega_{\mathrm{ref}}\left(\frac{f}{f_{\mathrm{ref}}}\right)^{\alpha_1} \bigg(1 + \left(\frac{f}{f_{\mathrm{break}}}\right)^{\frac{1}{\delta}}\bigg)^{\delta(\alpha_1 - \alpha_2)},
\end{equation}
\end{widetext}
consists of a power law spectrum with amplitude $\Omega_{\mathrm{ref}}$ and slope $\alpha_1$, which turns over at the break frequency $f_{\mathrm{break}}$ to a second power law with slope $\alpha_2$. The sharpness of the turnover is mediated by a smoothing parameter $\delta$, which is usually fixed to a fiducial parameter rather than inferred. The amplitude $\Omega_{\mathrm{ref}}$ is normalized such that it is the projected amplitude of the first power law at $f_{\mathrm{ref}}$ Hz. This model is a good approximation for the LMC DWD ASGWB established in \citet{rieck_stochastic_2024}.

\subsubsection{Analytic Foreground (Spectral)}\label{sec:blip2_models_analyticfg}
This model is a direct adaptation of the form given in \citet{cornish_galactic_2017a}, originally proposed by Babak as noted in \citep{cornish_galactic_2017a}:
\begin{widetext}
\begin{equation}
    S_h(f) = A f^{-7/3}e^{-f^{\alpha_{\mathrm{C17}}}-\beta f \sin(\kappa f)} \left[1 + \tanh(\gamma (f_k - f))\right] \mathrm{Hz}^{-1},
\end{equation}
\end{widetext}
which gives the Galactic foreground PSD as a function of an amplitude $A$ and a set of shape parameters $\{\alpha_{\mathrm{C17}},\beta,\kappa,\gamma,f_k\}$\footnote{$\alpha_{\mathrm{C17}}$ is given as $\alpha$ in \citet{cornish_galactic_2017a}; the ``C17" subscript included here is to disambiguate between this variable and the more general power law slope $\alpha$.} which evolve over time as more individual DWD systems are resolved out of the Galactic foreground; see \citet{cornish_galactic_2017a} for further details. As implemented in \blipp, this spectral model is recast in terms of $\Omega_{\mathrm{GW}}$:
\begin{widetext}
\begin{equation}
    \Omega_{\mathrm{GW}}(f)=\Omega_{\mathrm{ref}}\left(\frac{f}{f_{\mathrm{ref}}}\right)^{2/3}e^{-f^{\alpha_{\mathrm{C17}}}-\beta f \sin(\kappa f)}\left[1 + \tanh(\gamma (f_k - f))\right] 
\end{equation}
\end{widetext}
and adjusts the shape parameters based on the user's choice of observing duration.

\subsubsection{Analytic MW Satellite (Spatial)}\label{sec:blip2_spatial_satellite}
The Analytic MW Satellite submodel was initially implemented as a simplified model of the LMC, but its parameters can be modified to serve as an analog of any approximately spherical dwarf galaxy satellite of the MW.\footnote{i.e., any one of the nearby dwarf galaxies which orbit the MW: the Large and Small Magellanic Clouds, Sagittarius Dwarf, etc..} It is an appropriate model for any DWD ASGWB that would be better represented by a spatially extended distribution rather than a point source. It functions similarly to the Analytic Galaxy model in that it creates a parameterized 3D density distribution which is then projected onto the sky as seen by LISA. It creates a filled sphere of evenly spaced points centered at the location specified by its RA and DEC parameters, given in degrees and corresponding to standard ICRS right ascension and declination; at the specified distance $d$ from the Solar System Barycenter given in kpc. The radius of the sphere is set by the parameter $r$, given in kpc. GW power is evenly distributed among these points and projected onto a Healpix skymap.

\subsubsection{Point Source/Radiometer (Spatial)}\label{sec:blip2_spatial_radiometer}
The point source spatial submodel consists of a variable number of individual bright pixels on a Healpix \citep{gorski_healpix_2005,zonca_healpy:_2019} skymap; these can be contiguous or non-contiguous as desired. All pixels within a single point source submodel share a spectrum. The user can create multiple spectrally-independent point sources by instantiating several point source submodels, each containing a single source with its own individualized spectral submodel. When used for inference, this submodel effectively becomes a radiometer analysis \citep{ballmer_radiometer_2006b,mitra_gravitational_2008} and searches for stochastic GW power emanating from a particular sky direction or set of directions.

\subsubsection{Noise Spectral Model}\label{sec:blip2_spec_noise}
The current LISA instrumental noise model supported by \blipp\footnote{This noise model has been detailed in the context of \blip elsewhere \citep{banagiri_mapping_2021a,rieck_stochastic_2024,criswell_templated_2025,bloom_angular_2024a}, but is included here for completeness.}  is the two-parameter model given in \citet{amaro-seoane_laser_2017a}:
\begin{widetext}
\begin{equation}\label{eq:blip_noise_functions}
    \begin{split}
        S_p(f) &= N_p\left[1+\left(\frac{2\,\mathrm{mHz}}{f}\right)^4\right]\,\mathrm{Hz}^{-1},\\
        S_a(f) &= \left[1+\bigg(\frac{0.4\,\mathrm{mHz}}{f}\bigg)^2\right]\,\left[1+\bigg(\frac{f}{8\,\mathrm{mHz}}\bigg)^4\right] \frac{N_a}{(2\pi f)^4}\,\mathrm{Hz}^{-1},
    \end{split}
\end{equation}
\end{widetext}
where $S_p(f)$ and $S_a(f)$ describe the contributions due to position and acceleration noise, parameterized respectively by the noise amplitudes $N_p$ and $N_a$. This noise model is used for both noise simulation and inference. The actual LISA instrumental noise is expected to be significantly more complex than what is considered here. Detailed, realistic models of the LISA instrumental noise are under development \citep{littenberg_prototype_2023,bayle_unified_2023a,hartwig_stochastic_2023, pagone_noise_2024} and should be incorporated in future; however, such efforts are beyond the scope of this work.

\subsubsection{Spectral Models with Fixed Parameters}
For the extant models discussed above, it is often beneficial to fix (i.e., hold constant and not sample over) one or more parameters. This can be desirable for several reasons. From an astrophysical perspective, we expect (for example) the SOBBH SGWB to have a slope of $\alpha=2/3$; it may behoove us to search for a power law ISGWB in LISA with that slope. Similarly, DWD SGWB spectra are expected to exhibit a $2/3$ slope at low frequencies before the effects of discreteness and binary interactions become relevant. Provided the Universe conforms to our expectations, fixing one or more parameters to their physically-motivated values can make our search more efficacious. From a pure modelling perspective, one may wish to prescribe that certain parameters like the truncated power law's $f_{\mathrm{scale}}$ --- which have little astrophysical meaning and tend to not vary much within a given use case --- be fixed to some fiducial value. \blip's \submodel structure flexibly allows for the creation of a \submodel variant to any of these spectral models which fixes any number of model parameters. Some pertinent cases, like those with a fixed $\alpha=2/3$ slope, are hardcoded and available as out-of-the-box spectral {\tt submodel}s.

\subsection{Optimization}
Many \blip analyses previously required a prohibitive amount of computational resources and time to perform. For instance, some anisotropic analyses --- which must convolve the LISA responses with a model skymap while sampling --- could take several months to converge within the original \blip framework (assuming they did in fact converge). Even before the MCMC sampling step, computing the LISA responses for every time segment, across the full LISA frequency band, for every sky direction, and (in the spherical harmonic case) every spherical harmonic mode up to some truncation $\ell^a_{\mathrm{max}}$ to simulate a SGWB --- let alone multiple --- for the nominal LISA mission duration of 4 years was beyond the computational resources and time allocation available to the authors.

Many avenues have been pursued to push the limits of what \blip can accomplish with the computational power available. At a base level, more efficient trade-offs between memory usage and computation time have been implemented in the response function calculations, and the new modular, streamlined structure of \blipt alleviates many instances of repeated calculations. Beyond these basic coding improvements, more involved acceleration techniques have been implemented within {\tt BLIP}, each falling into one of three categories: parallelization, just-in-time compilation with JAX, and GPU-acceleration.

\subsubsection{Parallelization}
Parallelization has been implemented across several modules of the \blip codebase. The LISA response function calculations --- far and away the most computationally-taxing component of \blip --- are now fully parallelized along their frequency axis. This has reduced the required computation time for the time-dependent LISA response by over an order of magnitude for the most complex analyses considered in this work.\footnote{With an even greater reduction for less intensive analyses; total available RAM remains a fundamental limitation on the number of active parallel processes.} Additionally, each \submodel can be computed in parallel to the others,\footnote{(although this is mutually exclusive with the usually-superior response function parallelization --- the latter being an improvement on the former --- due to Python's multiprocessing limitations)} and parallel sampling routines have been implemented to work with the {\tt dynesty} and Numpyro samplers.

\subsubsection{Just-in-Time Compilation with JAX}
All parts of the \blip code that interact with its sampling routines (i.e., the \model and \submodel objects and everything attached to them) are now fully compatible with JAX. JAX \citep{bradbury_jax:_2018a} is a Python library that --- modulo several specific design requirements for user code\footnote{See \href{https://jax.readthedocs.io/en/latest/notebooks/Common_Gotchas_in_JAX.html}{the JAX documentation}.} --- serves as an approximately 1:1 swap in for Numpy \citep{harris_array_2020a} with several powerful additions. First among these is just-in-time compilation (JIT) of Python code. Python is not a compiled coding language; as such, it gains significant flexibility in exchange for reduced efficiency. JAX's JIT compiles code just as it is about to be executed; the resulting speed-up is usually negligible for one-time operations, but for functions that will be called repeatedly,\footnote{A pertinent example being the likelihood and prior given to a sampler.} the improvement can be dramatic. Second, JAX enables automatic differentiation, which allows the gradient to be taken of any JAX-able Python function; crucially, this includes the \blip likelihood, allowing its gradient with respect to its (potentially many) parameters to be evaluated and enabling the use of Hamiltonian Monte Carlo (HMC) sampling with \blip. Due to \blips new modular structure, any subsequent spatial or spectral models added to \blip in future will automatically be JAX-compliant. Finally, JAX is also designed to function seamlessly across both CPU and GPU settings.

\subsubsection{GPU-Accelerated Sampling with Numpyro}
These latter points lead to the final major speed improvement for \blip analyses: the implementation of GPU-accelerated HMC sampling with Numpyro \citep{phan_composable_2019a}. Numpyro is built with JAX in mind, using the latter's automatic differentiation to resolve the central challenge of HMC (i.e., needing to know the gradient of the likelihood) and leveraging JAX's innate GPU compatibility to do so at breathtaking speeds. On Nvidia A100 GPUs, convergence times for ISGWB and low-resolution spherical harmonic/pixel-basis ASGWB analyses for one year of LISA data were reduced from order days to weeks to order minutes to hours. As will be seen throughout the rest of this work, most analyses over the nominal LISA mission data duration of 4 years have been brought in reach, taking $\sim$a day on an Nvidia A100 and $\mathcal{O}$(tens of minutes --- a few hours) on an Nvidia GH200 superchip. The sole exception to this new computational horizon is high-resolution, parameterized anisotropic searches such as that of the spherical harmonic analysis at high $\ell^a_{\mathrm{max}}$. The final response functions for such analyses carry RAM requirements ranging from hundreds of GBs to a few TBs; a far cry from being able to fit within the available 80GB of RAM on a A100 GPU, or even the 141GB of an Nvidia H200 GPU. The GH200 superchip's shared memory presents promising solution to this limitation and can in principle enable GPU-acceleration of parameterized, high-resolution anisotropic analyses. At present, however, JAX does not yet support JIT with arrays distributed across shared memory; making use of this powerful resource will require further development that we leave for future work. Finally, computation of the time-dependent directional LISA response function and time-domain data has also been GPU-accelerated, with simulation of a multi-component, 4-yr dataset reduced to $\mathcal{O}$(hours); shorter and/or less-complex datasets can take as little as a few minutes for the entire data simulation process.

The advances highlighted here represent a factor of $\mathcal{O}(10^4) - \mathcal{O}(10^5)$ improvement in efficiency, and have achieved a major milestone in \blips development: the runtime of most \blip analyses is now computationally feasible in a Global Fit setting, fitting many times over within, e.g., the $\sim$week runtime of the similarly GPU-accelerated Erebor \citep{katz_efficient_2024}. The stage is set for inference of multiple isotropic and anisotropic SGWBs within the Global Fit itself.

\subsection{Utility Features}\label{sec:blip_utility}
In addition to the full code restructure, a number of utility and quality-of-life features have been added to \blipp. These include: 
\begin{enumerate}
    \item Checkpointing of all analyses at a user-defined interval to preserve progress in the event a long run is interrupted.
    \item The ability to resume sampling of chains from the final state of a completed run. This is useful for MCMC samplers like Numpyro that do not have the well-defined convergence criteria of nested sampling.
    \item The ability to produce and save a simulation for later analysis, as opposed to the previous requirement of a joint simulation + analysis run; this saves a not-insignificant amount of computational expense, especially for repeated analyses of the same dataset. Doing so stores not only the produced time-domain dataset, but also the entire \inj object, allowing \blip to carry forward all pertinent simulation information for evaluation of analysis results.
    \item All computed simulated and estimated spectra and skymaps used to create the final set of \blip plots are saved to a separate compressed file, the contents of which can be easily accessed outside of the \blip structure and its dependencies. This feature greatly eases the process of creating and reformatting plots of \blip results.
    \item Many other small aesthetic and ease-of-use improvements, such as: dynamic sizing of corner and spectral plots; the ability to specify a skymap color scheme for anisotropic recoveries; the ability to alias simulation {\tt submodel}s to those used in the analysis for purpose of plotting, carrying around ``true" (simulated) parameter values, etc.; and more.
\end{enumerate}

\begin{table*}
    \centering
    \begin{tabular}{|c|c|c|c|c|c|c|c|}
        \hline
         Component & Spectral & Parameters & Value & Spatial & Parameters & Values & Notes \\
         \hline
         \multirow{2}{6em}{\centering Noise} & \multirow{2}{6em}{\centering Instrumental Noise} & $N_a$ & $3.6\times10^{-49}$ Hz$^{-4}$ & \multirow{2}{4em}{\centering ---} & \multirow{2}{4em}{\centering ---} & \multirow{2}{4em}{\centering ---} & \multirow{2}{10em}{\centering Per \citet{amaro-seoane_laser_2017a}.} \\
         &  & $N_p$ & $9\times10^{-42}$&  &  &  & \\
         \hline
         \multirow{4}{6em}{\centering MW} &  & $\Omega_{\rm ref}$ & $2.197\times10^{-9}$ &  &  &  & \multirow{4}{10em}{\centering Thin disk MW model of \citet{breivik_constraining_2020c}.}\\
         & Tanh-truncated & $\alpha$ & 0.667 & Analytic & $r_h$ & 2.9 kpc & \\
         & Power Law & $f_{\rm cut}$ & 2.08 mHz & Galaxy & $z_h$ & 0.3 kpc & \\
         &  & $f_{\rm scale}$ & 0.5 mHz &  &  &  & \\
         \hline
         \multirow{5}{6em}{\centering LMC} &  & $\Omega_{\rm ref}$ & $7.6\times10^{-12}$ &  & $r$ & 2.15 kpc & \multirow{5}{10em}{\centering Based on LMC spectrum of \citet{rieck_stochastic_2024}.} \\
         & \multirow{3}{6em}{\centering Broken Power Law} & $\alpha_1$ & 0.667 & \multirow{3}{4em}{\centering Analytic Satellite} & $d$  & 50 kpc & \\
         & & $\alpha_2$ & 2.65 & & RA & 5h23m34s & \\
         &  & $f_{\rm break}$ & 3.83 mHz &  & DEC & -69\textdegree45'22'' & \\
         &  & $\delta$ & 0.1 &  & &  & \\
         \hline
         \multirow{2}{6em}{\centering SOBBH} & \multirow{2}{6em}{\centering Power Law} & $\Omega_{\rm ref}$ & $2.11\times10^{-12}$ & \multirow{2}{4em}{\centering Isotropic} & \multirow{2}{4em}{\centering ---} & \multirow{2}{4em}{\centering ---} & \multirow{2}{10em}{\centering Upper range estimate of \citet{babak_stochastic_2023a}.}\\
         &  & $\alpha$ & 0.667 &  &  &  & \\
         \hline
         \multirow{2}{6em}{\centering CGWB} & \multirow{2}{6em}{\centering Power Law} & $\Omega_{\rm ref}$ & $1.0\times10^{-13}$ & \multirow{2}{4em}{\centering Isotropic} & \multirow{2}{4em}{\centering ---} & \multirow{2}{4em}{\centering ---} & \\
         &  & $\alpha$ & 1.0 &  &  &  & \\
         \hline
    \end{tabular}
    \caption{Summary of simulation components. }
    \label{tab:simulations}
\end{table*}

\begin{table*}
    \centering
    \begin{tabular}{|c|c|c|c|c|c|c|}
        \hline
         Submodel & Spectral & Parameters & Prior/Fixed Value & Spatial & Parameters & Template Values \\
         \hline
         \multirow{2}{6em}{\centering Noise} & \multirow{2}{6em}{\centering Instrumental Noise} & $\log_{10}N_a$ & $\mathcal{U}(-51,-46)$ & \multirow{2}{4em}{\centering ---} & \multirow{2}{4em}{\centering ---} & \multirow{2}{4em}{\centering ---} \\
         &  & $\log_{10}N_p$ & $\mathcal{U}(-44,-39)$&  &  &  \\
         \hline
         \multirow{4}{6em}{\centering MW} &  & $\log_{10}\Omega_{\rm ref}$ & $\mathcal{U}(-10,-7)$ &  &  &  \\
         & Tanh-truncated & $\alpha$ & $\mathcal{U}(0,2)$ & Analytic & $r_h$ & 2.9 kpc \\
         & Power Law & $f_{\rm cut}$ & $\mathcal{U}(0.8,4)$ mHz & Galaxy & $z_h$ & 0.3 kpc \\
         &  & $f_{\rm scale}$ & $\mathcal{U}(0.1,10)$ mHz &  &  & \\
         \hline
         \multirow{5}{6em}{\centering LMC} &  & $\log_{10}\Omega_{\rm ref}$ & $\mathcal{U}(-9,-12)$ &  & $r$ & 2.15 kpc \\
         & \multirow{3}{6em}{\centering Broken Power Law} & $\alpha_1$ & $\mathcal{U}(0,2)$ & \multirow{3}{4em}{\centering Analytic Satellite} & $d$  & 50 kpc \\
         & & $\alpha_2$ & $\mathcal{U}(0,4)$ & & RA & 5h23m34s \\
         &  & $f_{\rm break}$ & $\mathcal{U}(1,10)$ mHz &  & DEC & -69\textdegree45'22'' \\
         &  & $\delta$ & 0.1 &  & &  \\
         \hline
         \multirow{2}{6em}{\centering SOBBH} & \multirow{2}{6em}{\centering Power Law} & $\log_{10}\Omega_{\rm ref}$ & $\mathcal{U}(-11,-13)$ & \multirow{2}{4em}{\centering Isotropic} & \multirow{2}{4em}{\centering ---} & \multirow{2}{4em}{\centering ---} \\
         &  & $\alpha$ & 0.667 &  &  &  \\
         \hline
         \multirow{2}{6em}{\centering CGWB} & \multirow{2}{6em}{\centering Power Law} & $\log_{10}\Omega_{\rm ref}$ & $\mathcal{U}(-21,9)$ & \multirow{2}{4em}{\centering Isotropic} & \multirow{2}{4em}{\centering ---} & \multirow{2}{4em}{\centering ---} \\
         &  & $\alpha$ & $1$ &  &  &  \\
         \hline
    \end{tabular}
    \caption{Summary of priors and model choices.}
    \label{tab:priors}
\end{table*}

\section{Simulations}\label{sec:sims}
The modular framework described in \S\ref{sec:blip2} also enables simulation of LISA datasets containing --- in principle --- any number and combination of isotropic and anisotropic SGWBs. Within \blipp, these are referred to as simulation (or injection) ``components''. Realistically, the number of components is of course limited somewhat by scaling computational expense, as well as our expectations of what will actually be present in LISA data. From a computational perspective, the primary limitation is the need to compute the full time-dependent directional LISA response functions $\mathcal{R}_{IJ}(f,t,\mathbf{n})$ for every simulated SGWB. At appreciable timescales and high fidelity of the simulated Healpix skymaps, calculating the LISA response carries with it the vast majority of the computational expense of a given simulation. To mitigate this, \blip dynamically shares information between individual simulation components to avoid repeated calculations and ensure that only the necessary calculations are performed. For example, if one wishes to simulate two point source SGWBs, each comprising a single pixel on the sky, \blip will only compute the underlying satellite orbits, transfer functions, etc. once, and will innately limit the directional calculations to the two pixels with power (as the remaining pixels have zero power and will sum to zero when convolving the directional LISA response functions with the SGWB skymaps; see Eqn.~\eqref{eq:si_skymap_convolution}). At present, \blip is able to simulate multi-SGWB datasets spanning a reasonable expectation of what will be present within LISA data\footnote{With the exception of a potential non-stationary, non-Gaussian SGWB from extreme mass-ratio inspirals \citep[e.g.,][]{buscicchio_test_2024a,xuan_stochastic_2024,naoz_enhanced_2023b}, as \blip does assume both Gaussianity and stationarity; see further discussion in \S\ref{sec:discussion_conclusions}.} for a duration of 4 years and with a Healpix {\tt nside} (pixel resolution) of 16. This capability is demonstrated as follows.

\subsection{Simulation Components}
We use the BLIP 2.0 framework to simulate several multi-SGWB datasets for use as test cases. These consist of the following components; for each, see \S\ref{sec:blip2_models}, Table~\ref{tab:models}, and references therein for detailed descriptions of the underlying spectral and spatial models. Model parameter values for the simulated SGWBs discussed here are given in Table~\ref{tab:simulations}.

\subsubsection{Simple Galactic Foreground (``MW'')}
We simulate a simple Galactic foreground, combining the smooth Tanh-truncated Power Law spectrum with a parametric model for the Galactic spatial density distribution (Analytic Galaxy) consisting of a simple bulge and exponential disk. The simulated Galaxy has radial and vertical scale heights of $r_h=2.9$ kpc and $z_h = 0.3$ kpc, respectively (the ``thin disk'' model of \citep{breivik_constraining_2020c}, based on previous models of \citet{binney_photometric_1997, bissantz_spiral_2002, mcmillan_mass_2011b}). \blip is capable of simulating and modeling more-realistic Galactic foregrounds produced from binary population synthesis catalogues (e.g., {\tt popmap} in Table~\ref{tab:models} as discussed in \citep{criswell_templated_2025,rieck_stochastic_2024}). However, performing detailed spectral separation between a realistic Galactic foreground and underlying SGWBs will comprise a significant effort beyond the scope of this work. A simple Galactic foreground better suits the current purpose of demonstrating \blips simulation and analysis capabilities --- as the simulated and inferred parameter values can be directly compared --- but we will explore the challenge of spectral separation with a realistic foreground in future work (see \S\ref{sec:discussion_conclusions}).

\subsubsection{Isotropic SOBBH SGWB (``SOBBH'')}
We simulate the SGWB from extragalactic stellar-origin binaries as an isotropic signal with a power-law slope of $\alpha=2/3$ and the upper range amplitude given in \citet{babak_stochastic_2023a}. This latter work provides projections of the LISA SOBBH SGWB based on the population of LVK-observed compact binary mergers.

\subsubsection{Simple LMC (``LMC'')}
We simulate a simple version of the anisotropic SGWB from white dwarf binaries in the LMC \citep{rieck_stochastic_2024}. This is an analytic simulation of the LMC SGWB for the same reasons discussed above with respect to the Galactic foreground. We simulate the LMC SGWB spectrum as a broken power law, due to its notable turnover but lack of rapid truncation; see, e.g., Figure 2 of \citet{rieck_stochastic_2024}. The LMC spatial distribution is simulated as a uniform-density sphere with the observed radius of the LMC.

\subsubsection{Cosmological SGWB (``CGWB'')}
We additionally simulate a generic cosmological SGWB. It is isotropic, lower in amplitude than the SOBBH SGWB, and possesses a flat power spectral density --- representative of a generic scale-invariant inflationary background. Its primary purpose is to demonstrate \blips ability to place upper limits on such a signal in the presence of a variety of higher-amplitude isotropic and anisotropic SGWBs.



\subsection{Simulated Datasets}
We consider two example datasets, each consisting of combinations of the SGWB components described above. Each dataset spans 4 years, corresponding to $1.262\times10^8$ s of simulated data. Anisotropy is simulated using Healpix skymaps at an {\tt nside} (pixel resolution) of 16. The reference frequency used throughout is $f_{\rm ref} = 1$ mHz. We assume Gaussian noise of the spectral form given in \S\ref{sec:blip2_spec_noise} with parameters $N_a$ and $N_p$ as given in \citet{amaro-seoane_laser_2017a} (see Table~\ref{tab:simulations}). We simulate time-domain LISA data by splicing together inverse Fourier transforms of the frequency-domain Gaussian contribution from each simulation component, convolved with the time-dependent directional LISA response across each considered TDI channel, on overlapping, Hann-windowed splice segments of duration $10^{-4}$ s. Further details can be found in \citet{banagiri_mapping_2021a}.

The first dataset, ``MW + LMC + SOBBH'' combines a set of likely SGWB signals: the anisotropic Galactic foreground, the highly localized ASGWB from white dwarf binaries in the LMC, and the ISGWB from extragalactic stellar-origin binaries (SOBBH). The second dataset ``MW + LMC + SOBBH + CGWB'' additionally includes an low-amplitude isotropic cosmological background to investigate our ability to place upper limits on an underlying SGWB signal of cosmological origin in the presence of multiple foreground signals. The data simulation process is fully GPU-accelerated, and takes $\sim 5$ hours to calculate the full $3\times3\times$ frequency $\times$ time $\times$ sky direction LISA response tensor and simulate time-domain data for a multi-component, 4-year dataset at the chosen skymap resolution on a single Nvidia GH200.

\section{Results}\label{sec:results}


\begin{figure*}
    \centering
    \includegraphics[width=1\linewidth]{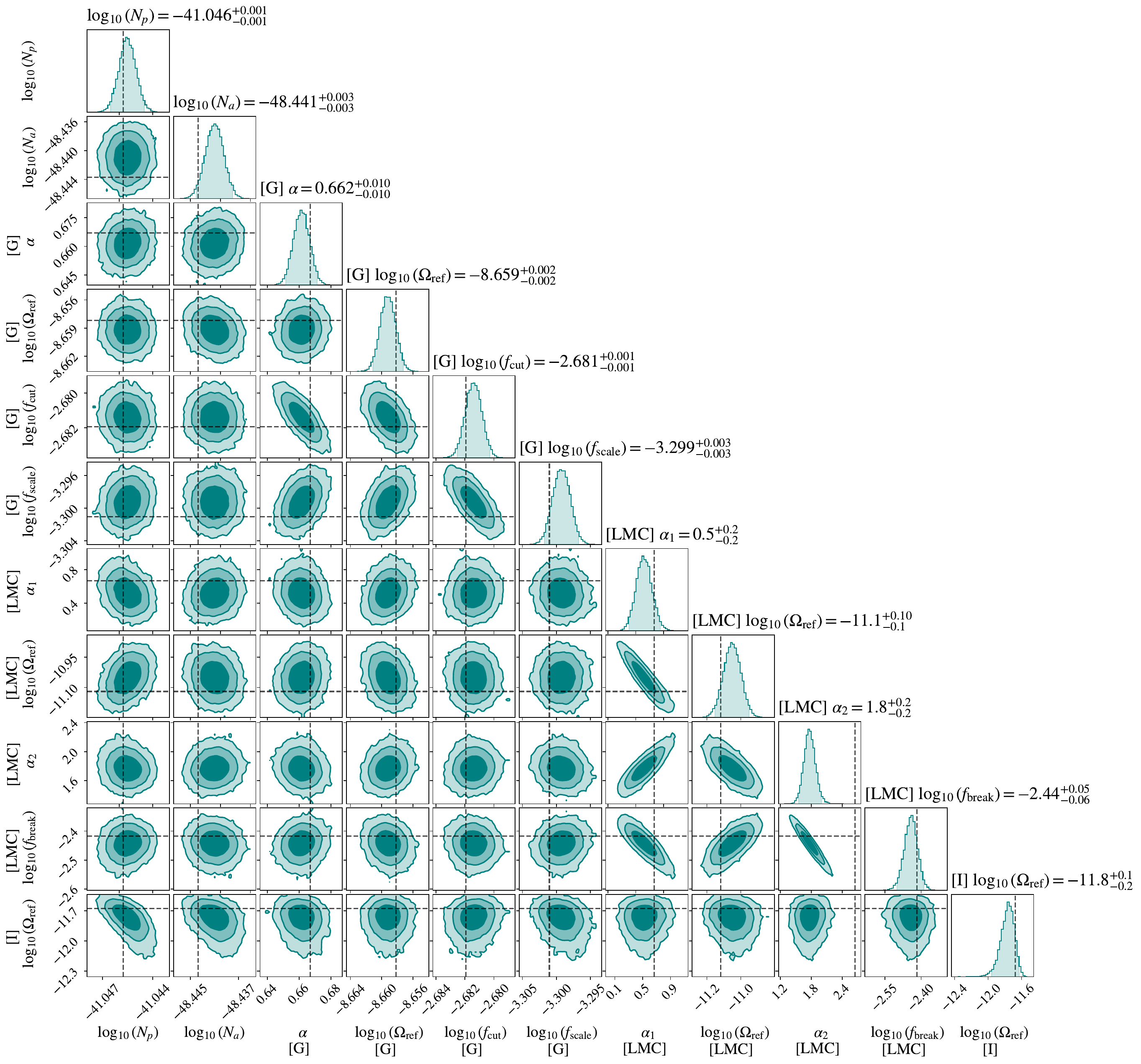}
    \caption{Corner plot showing the posterior samples for each inferred parameter of the three-signal MW+LMC+SOBBH model. Quoted bounds are the 95\% credible intervals, corresponding to the shaded region of the 1D marginal distributions. The 2D contours show the 68\%, 95\%, and 99.7\% credible levels.}
    \label{fig:three_signal_corner}
\end{figure*}

\begin{figure*}
    \centering
    \includegraphics[width=0.65\linewidth]{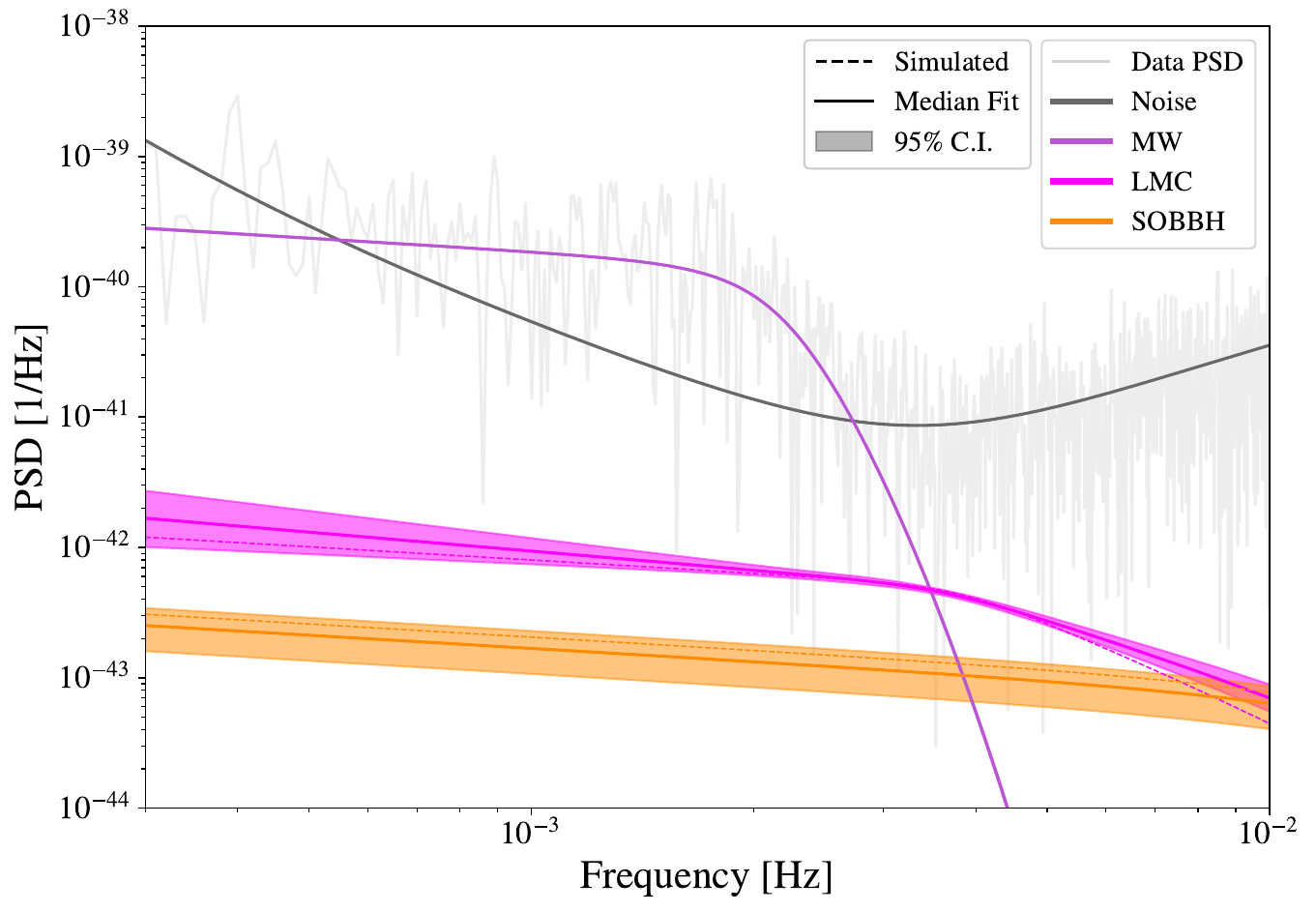}
    \caption{Simulated and inferred spectral distributions for the three-signal MW + LMC + SOBBH dataset. Dashed lines show the simulated spectra; solid lines and shaded regions show the median spectral fit and 95\% credible interval of the recovered spectra models. Note that due to its high amplitude, the 95\% C.I. region of the Galactic foreground spectral fit is much smaller than that of the other signals and --- while present in the plot --- is difficult to discern by eye. A single-segment data PSD is shown in light grey; note that the variance of the simulated data is greatly reduced when averaged over the $\mathcal{O}(10^3)$ segments in a 4 year dataset. Each signal is well-recovered, with the exception of some spectral mixing between the LMC and SOBBH SGWBs at high frequencies, resulting in overestimation of the second part of the broken power law.}
    \label{fig:three_signal_spectra}
\end{figure*}

\begin{figure*}
    \centering
    \includegraphics[width=0.65\linewidth]{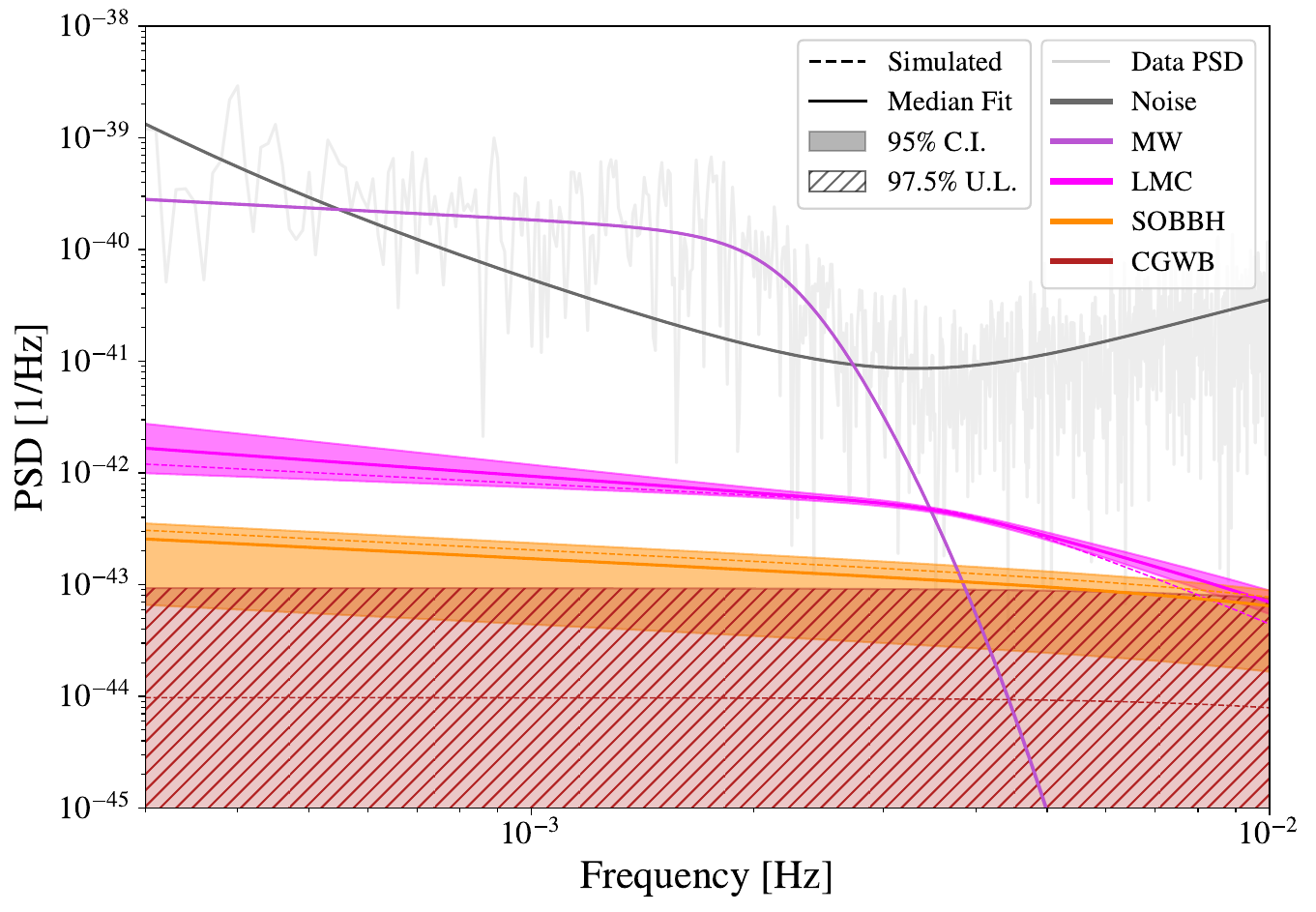}
    \caption{Simulated and inferred spectral distributions for the four-signal MW + LMC + SOBBH + CGWB dataset. Dashed lines show the simulated spectra; solid lines and shaded regions show the median spectral fit and 95\% credible interval of the recovered spectra models. Note that due to its high amplitude, the 95\% C.I. region of the Galactic foreground spectral fit is much smaller than that of the other signals and --- while present in the plot --- is difficult to discern by eye. A single-segment data PSD is shown in light grey; note that the frequency-scatter of the simulated data is greatly reduced when averaged over the $\mathcal{O}(10^3)$ segments in a 4 year dataset. The astrophysical SGWB spectra are well-recovered, with the exception of some spectral mixing between the LMC and SOBBH SGWBs at high frequencies, resulting in overestimation of the second part of the broken power law. An upper limit is placed on the CGWB in the presence of astrophysical foregrounds.}
    \label{fig:four_signal_spectra}
\end{figure*}

\subsection{MW + LMC + SOBBH}\label{sec:three_signal_analysis}
We analyze the 3-component dataset as follows. We use the Tanh-truncated Power Law spectral model for the MW spectrum, and the Analytic Galaxy pixel-basis template for the MW spatial model \citep[see Table~\ref{tab:models} and ][]{criswell_templated_2025}. The MW model Galaxy morphology is matched to that of the simulation, i.e. with scale height parameters set to $r_h=2.9$ kpc and $z_h = 0.3$ kpc; we can make this assumption as the spatial distribution of the DWD population will be well-measured by the resolved DWDs \citep[e.g.,][]{wilhelm_milky_2020}. The LMC spectrum is modelled with a broken power law; the LMC spatial distribution is modelled with the Analytic Satellite pixel-basis template (see \S\ref{sec:blip2_spatial_satellite}). We use a power law ISGWB to model the SOBBH ISGWB, with a fixed slope of $\alpha=2/3$ to target the search for the expected shape of the SOBBH spectrum. We use the noise model described in \S\ref{sec:blip2_spec_noise}. The priors for each search are astrophysically-motivated so as to target their respective signal; these priors are given in Table~\ref{tab:priors}. The inferred posterior distributions are shown in Fig.~\ref{fig:three_signal_corner}, and the recovered spectra are shown in Fig.~\ref{fig:three_signal_spectra}. 

The Galactic foreground spectrum is precisely and accurately recovered; this is unsurprising given its prominence, the matching functional form of the simulated spectrum, and the strong prior enforced by the fixed-template spatial analysis. However, the agreement of the inferred parameters with the simulated values as seen in Fig.~\ref{fig:three_signal_corner} provides an important consistency test. That being said, the recoveries of the lower-amplitude signals are more interesting. First, the LMC is overall recovered well. The templated anisotropic search allows for recovery of this more-complex spectral model compared to the power law + spherical harmonic analysis of \citet{rieck_stochastic_2024}, and recovery of the LMC spectrum's low-frequency amplitude and slope is much improved compared to that initial study. Its (log) amplitude is measured to within 10\%, indicating an intriguing potential for comparison of the LMC and Galactic SGWB spectra --- and therefore the masses, number, and typical distances of their underlying DWD populations. Adding to this potential is the precise and accurate inference of the turn-over frequencies for both the MW and LMC spectra. For realistic signals, this phenomenological quantity is driven by both characteristics of these astrophysical populations and detector-dependent effects (vis-\`a-vis resolving out individual DWDs). If these effects can be separated, the different turnovers of the LMC and MW spectra may provide additional angles through which to investigate the underlying astrophysics of their separate DWD populations.

Conversely, we see significant spectral mixing and a corresponding bias in the recovery of the second LMC power law below the turnover frequency $f_{\rm break}$ (i.e., for $f\gtrsim 5$ mHz). As can be seen in Fig.~\ref{fig:three_signal_spectra}, the spectrum is accurately inferred around the knee, until the noise level raises substantially and the amplitudes of the LMC and SOBBH signals become comparable. Further investigation is needed to determine if this bias presents a serious issue at higher frequencies, especially when considering realistic spectra from actual (simulated) DWD populations which are generally much more complex at higher frequencies. We note this as a direction for future work. 

Finally, the simulated SOBBH SGWB log amplitude is inferred to be $\log_{10}\Omega_{\rm GW} = -11.8^{+0.1}_{-0.2}$ (95\% C.I.), in agreement with the simulated value of $-11.68$. Even when considering multiple astrophysical foregrounds, this constraint improves upon the expectations derived from observations of compact binary mergers as of GWTC-3; c.f. the range calculated in \citet{babak_stochastic_2023a} of $-12.0^{+0.3}_{-0.5}$ (converted here to units of $\log_{10}\Omega_{\rm GW}(1\ \mathrm{mHz})$ for the sake of comparison). Continued LVK observations will of course continue to narrow these predictions; agreement (or lack thereof) between LISA and LVK observations of these systems will provide an exciting window into the evolution of stellar-origin compact binaries. Conversely, projections from LVK-observed mergers \citep[e.g.,][]{abbott_upper_2021a} are necessarily model-dependent; combining information across both LISA and LVK data is a promising route for future multi-band science. It is worth noting that both these and previous constraints on the SOBBH, LMC, and MW spectra depend on accurate characterization of the LISA instrumental noise. Here, we assume knowledge of the functional form of the noise spectrum; this will not be the case when LISA flies, and as such these estimates can be expected to degrade somewhat under realistic conditions \citep{hartwig_stochastic_2023,muratore_impact_2023}.

\begin{figure*}
    \centering
    \includegraphics[width=0.65\linewidth]{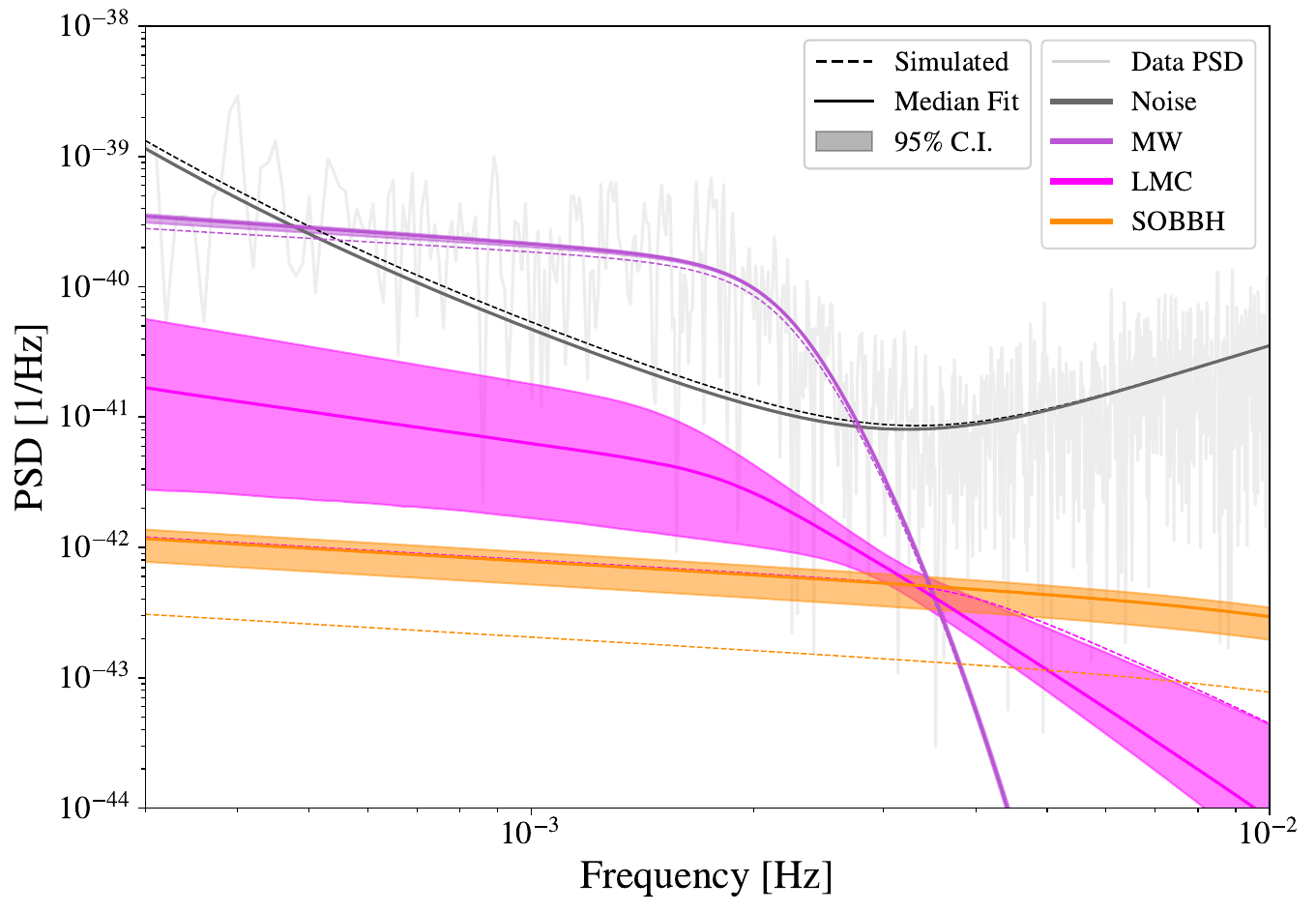}
    \caption{Simulated and inferred spectral distributions for the three-signal MW + LMC + SOBBH dataset, where the anisotropic nature of the MW and LMC have been neglected. Dashed lines show the simulated spectra; solid lines and shaded regions show the median spectral fit and 95\% credible interval of the recovered spectra models. A single-segment data PSD is shown in light grey; note that the frequency-scatter of the simulated data is greatly reduced when averaged over the $\mathcal{O}(10^3)$ segments in a 4 year dataset. Significant spectral mixing occurs, and mismodelling the spatial distribution of these signals results in severely biased results.}
    \label{fig:three_signal_alliso}
\end{figure*}

\subsection{MW + LMC + SOBBH + CGWB}\label{sec:four_signal_analysis}
We now analyze the four-signal MW + LMC + SOBBH + CGWB dataset. We make identical modelling choices as in the previous analysis, with the inclusion of an additional isotropic power law model. We perform a targeted search for a power law of fixed spectral index $\alpha=1$ with a broad prior on the amplitude; all priors used are given in Table~\ref{tab:priors}. We place an upper limit on the amplitude of the simulated CGWB in the presence of astrophysical foregrounds of $\log_{10}\Omega_{\rm GW}(1 {\rm\ mHz}) \leq -12.0$ ($97.5\%$ credible upper bound). This is roughly consistent with the result of e.g., \citet{boileau_spectral_2021d}, who are able to recover a CGWB of amplitude $\log_{10}\Omega_{\rm GW}(1 {\rm\ mHz}) =  -12.097$ (although the comparisons are not precisely 1:1 due to different data durations, slopes, amplitudes of astrophysical foregrounds, inclusion of the LMC, and so on). However, detailed consideration of CGWB detection thresholds in the presence of multiple astrophysical foregrounds warrants a dedicated investigation beyond the scope of this work. 

It is worth noting that due to the isotropic nature of both signals, the inclusion of the CGWB increases the uncertainty of the inferred SOBBH amplitude. While the uncertainty of the MW and LMC estimates are unchanged by inclusion of the underlying CGWB, the SOBBH estimate degrades slightly: from $\Omega_{\rm GW}(1 {\rm\ mHz}) = -11.8^{+0.1}_{-0.2}$ (95\% C.I.) in the three-signal case to $\Omega_{\rm GW}(1 {\rm\ mHz}) = -11.8^{+0.1}_{-0.6}$ (95\% C.I.) in the four-signal case. Power is able to be freely traded between the two isotropic signal models, and the SOBBH inference suffers as a result. However, it is important to note that while the inclusion of the CGWB increases the uncertainty of the SOBBH amplitude estimate, it does not induce a bias.

\subsection{Effect of Neglecting Anisotropy in Spectral Separation}\label{sec:anisotropy_neglect}
We additionally, consider the impact of neglecting the anisotropy of the MW and LMC signals. We repeat the MW+LMC+SOBBH analysis of \S\ref{sec:three_signal_analysis} keeping all aspects of the analysis identical save that the MW, LMC, and SOBBH SGWBs are all modelled with isotropic spatial distributions. The resulting inferred spectra are shown in Fig.~\ref{fig:three_signal_alliso}. Failing to model the intrinsic anisotropy greatly increases spectral mixing and induces significant systematic biases in the inferred spectra. The MW, LMC, and SOBBH amplitudes are overestimated by 2\%, 830\%, and 376\%, respectively, with each inferred amplitude completely excluding the corresponding simulated value. These biases arise from a combination of unmodeled low-frequency power due to the anisotropy-induced time-modulation and absorption of instrumental noise into the SGWB power, as the acceleration noise is underestimated by 86\% whereas the position noise estimate is unbiased (the latter being dominant at higher frequencies). The severity of these effects dramatically underlines the crucial importance of anisotropy for accurate SGWB spectral separation with LISA.

\section{Discussion and Conclusions}\label{sec:discussion_conclusions}
We present {\tt BLIP 2.0}, a flexible framework for Bayesian inference of multiple isotropic and anisotropic SGWBs in LISA. Its modular structure allows for assembly of inference models composed of arbitrary numbers and morphologies of SGWB signals, up to computational limitations. The same structure likewise enables simulation of multi-component LISA SGWB datasets. Accounting for signal morphology enhances spectral separation of multiple signals, and allows for spectro-spatial separation of multiple ASGWBs. This is of particular importance for LISA due to the prominent anisotropic Galactic foreground and other local-Universe ASGWBs such as the LMC. We present two multi-component simulated datasets and Bayesian SGWB analyses thereof to demonstrate the capabilities of the framework: the first consisting of the Galactic foreground, the LMC SGWB, and the SOBBH ISGWB, and the second additionally including an underlying cosmological SGWB. We briefly demonstrate the adverse effects of neglecting anisotropy when attempting spectral separation of LISA SGWBs. The {\tt BLIP 2.0} framework is the first of its kind, and greatly expands our analysis capability for ASGWB analyses and spectro-spatial separation of multiple isotropic and anisotropic SGWBs with LISA. Additionally, its Bayesian approach and modular infrastructure are well-suited for integration into Bayesian Global Fits.

There remain some limitations of the code. It currently operates on a series of short-time Fourier transforms, which are less robust to non-stationary noise and data gaps; implementation of the time-frequency basis \citep{cornish_timefrequency_2020} is a logical next step. Doing so would also enable exploration with \blip of non-stationary SGWBs like that expected from an EMRI SGWB, and aid in Global Fit integration. Additionally, the noise modelling used in the demonstrations presented here is quite simple compared to our expectations for the LISA instrumental noise; implementation of more advanced noise modelling is warranted, and will be supported by \blips modular framework.

Looking forward, it will be important to consider the case of non-idealized spectra. The Galactic foreground will not have a smooth spectrum, and will include features from both astrophysical populations \citep[e.g.,][]{scaringi_cataclysmic_2023} and potential non-Gaussianities from individual subthreshold DWDs \citep{rosati_prototype_2024}. This will complicate spectral separation, especially for other DWD-driven SGWBs like that of the LMC. In this work we were able to perform spectral separation between simple models of the MW and LMC; additional work will be needed to approach the realistic case of spectro-spatial separation between two DWD populations' SGWBs. This is an exciting direction of future work that will be greatly aided by \blips existing spatial models and straightforward support for new spectral spatial models. The results presented here indicate strong potential for LISA's ability to characterize important morphological aspects of the MW and LMC spectra, opening a path towards astrophysical comparison between these two local DWD populations.

The challenge of properly modelling the stochastic spectra of the MW and LMC arises from their very nature as astrophysical populations. Simple phenomenological models like those employed in this work, despite accurately describing the overall spectral shape of the MW and LMC, fail to incorporate the detailed effects of bin-to-bin Poisson fluctuations in the number and amplitude of the contingent unresolved DWDs. Moreover, the phenomenological truncation mechanisms are intrinsically unable to account for the departures from true stochasticity as source discreteness becomes relevant \citep[e.g.,][]{sesana_stochastic_2008}. On top of this, such models are difficult to interpret from an astrophysical perspective, as their shape can be just as influenced by detector-dependent resolution of individual binaries as they are by the underlying astrophysical population. Taken together, these factors strongly motivate development spectral models for the Galactic foreground (as well as the LMC) which directly arise from treatment of the unresolved astrophysical population. Not only would such models allow for more flexible and accurate characterization of realistic DWD-driven SGWBs, but also for astrophysical population inference from the Galactic foreground and --- potentially --- the LMC SGWB. In the case of the Galactic foreground, joint population inference with the resolved binaries may even allow for stronger constraints on the Galactic foreground (c.f. a resolved + unresolved population model for next-generation terrestrial detectors \citep{zhong_twostep_2025}), thereby improving almost every LISA science case. 

Finally, \blips modular infrastructure provides a strong foundation for transdimensional multi-SGWB modelling, wherein new submodels are proposed and removed, and the data itself can inform the choice of both spectral and spatial modelling. Work is underway towards this end, beginning with the implementation in \blip of {\tt Eryn} \citep{karnesis_eryn_2023, katz_mikekatz04_2023}, the transdimensional GPU-accelerated sampler used in the Erebor global fit \citep{katz_efficient_2024}.

\section*{Data Availability}
\blip is open source; the primary repository can be found at \hyperlink{https://github.com/sharanbngr/blip}{https://github.com/sharanbngr/blip}. The specific version of the code used in this study can be found within the corresponding BLIP release on Zenodo \citep{criswell_criswellalexander_2025}.  Simulated data and posterior samples for all analyses are available on Zenodo \citep{criswell_datasets_2025}.




\section*{Acknowledgments}
Portions of this manuscript were adapted from the doctoral dissertation of AWC, \textit{Astrophysical Inferences from Multimessenger Ensembles}. AWC acknowledges support by NSF grant no. 2125764 and NASA grant 90NSSC19K0318. S.B. acknowledges support from NSF grant PHY-2207945 and from Australian Research Council (ARC) grant CE230100016. The authors are members of the NANOGrav collaboration, which receives support from NSF Physics Frontiers Center award number 1430284 and 2020265. SRT acknowledges support from an NSF CAREER no. 2146016, NSF AST-2007993, and NSF AST-2307719. SRT also acknowledges support from a Chancellor's Faculty Fellowship from Vanderbilt University. The authors acknowledge the use of computing resources provided by the Minnesota Supercomputing Institute at the University of Minnesota. This work leveraged the resources provided by the Vanderbilt Advanced Computing Center for Research and Education (ACCRE), a collaboratory operated by and for Vanderbilt faculty.

%


\section*{Software}
{\tt BLIP} \citep[this work; ][]{banagiri_mapping_2021a,criswell_templated_2025}, {\tt numpy} \citep{harris_array_2020a}, {\tt scipy} \citep{virtanen_scipy_2020a}, {\tt ChainConsumer} \citep{hinton_chainconsumer_2016}, {\tt astropy} \citep{robitaille_astropy:_2013,theastropycollaboration_the_2018,astropycollaboration_the_2022a}, {\tt healpy} \citep{gorski_healpix_2005}, JAX \citep{bradbury_jax:_2018a}, Numpyro \citep{phan_composable_2019a}, and {\tt Matplotlib} \citep{hunter_matplotlib:_2007}.



\bibliography{LISA_Multi_SGWB,Code_Packages.bib}{}
\bibliographystyle{aasjournal}



\end{document}